\documentclass[pdflatex,sn-mathphys]{sn-jnl}% Math and Physical Sciences Reference Style
\usepackage{enumerate}
\usepackage{amsmath,amsthm}
\usepackage{amssymb, latexsym}
\usepackage[mathscr]{euscript}
\usepackage{natbib}
\usepackage{hyperref}

\jyear{2021}%

\theoremstyle{thmstyleone}%
\theoremstyle{thmstyletwo}%
\newtheorem{remark}{Remark}%

\theoremstyle{thmstylethree}%

\begin{document}

\title[Size does not matter]{Sometimes size does not matter}

\author*[1]{\fnm{Daniel Andr\'es} \sur{D\'iaz-Pach\'on}}\email{Ddiaz3@miami.edu}

\author[2]{\fnm{Ola} \sur{H\"ossjer}}\email{ola@math.su.se}

\author[3]{\fnm{Robert J.} \sur{Marks} \sfx{II}}\email{Robert\_Marks@Baylor.edu}

\affil*[1]{\orgdiv{Division of Biostatistics}, \orgname{University of Miami}, \orgaddress{\street{1120 NW 14 ST Suite 1057}, \city{Miami}, \postcode{33136}, \state{Florida}, \country{United States of America}}}

\affil[2]{\orgdiv{Department of Mathematics}, \orgname{Stockholm University}, \orgaddress{\street{Roslagsv\"agen 101, Kr\"aftriket, hus 6 Room 318}, \state{Stockholm}, \country{Sweden}}}

\affil[3]{\orgdiv{Dept. of Electrical \& Computer Engineering}, \orgname{Baylor University}, \orgaddress{\street{One Bear Place}, \city{Waco}, \postcode{76798}, \state{Texas}, \country{United States of America}}}


\abstract{Cosmological fine-tuning has traditionally been associated with the narrowness of the intervals in which the parameters of the physical models must be located to make life possible. A more thorough approach focuses on the probability of the interval, not on its size. Most attempts to measure the probability of the life-permitting interval for a given parameter rely on a Bayesian statistical approach for which the prior distribution of the parameter is uniform. However, the parameters in these models often take values in spaces of infinite size, so that a uniformity assumption is not possible. This is known as the normalization problem. This paper explains a  framework to measure tuning that, among others, deals with normalization, assuming that the prior distribution belongs to a class of maximum entropy (maxent) distributions. By analyzing an upper bound of the tuning probability for this class of distributions the method solves the so-called weak anthropic principle, and offer a solution, at least in this context, to the well-known lack of invariance of maxent distributions. The implication of this approach is that, since all mathematical models need parameters, tuning is not only a question of natural science, but also a problem of mathematical modeling. Cosmological tuning is thus a particular instantiation of a more general scenario. Therefore, whenever a mathematical model is used to describe nature, not only in physics but in all of science, tuning is present. And the question of whether the tuning is fine or coarse for a given parameter --- if the interval in which the parameter is located has low or high probability, respectively --- depends crucially not only on the interval but also on the assumed class of prior distributions. Novel upper bounds for tuning probabilities are presented.}




\maketitle

\section{Introduction}\label{Sec:Intro}

Cosmological fine-tuning is the idea that, in order for life to exist, the constants of nature must belong to intervals of very low probability. This definition, good for mnemotechnical purposes, can be made more rigorous by considering not only the constants individually, but also multiple ratios between them, and boundary conditions of the universe. For instance, the ratio of  the strong nuclear force to the weak nuclear force seems to be fine-tuned for life \citep[Ch.~4]{LewisBarnes2016}, or the difference between masses of the two lightest quarks (which corresponds to a ratio between their exponentials) also seems fine-tuned \citep{CarrRees1979}.  Accordingly, the fine-tuning problem seems to be divided into two stages, given that we know which are the relevant constants: 

\begin{itemize}
	\item[\textbf{(i)}] Finding the interval $I \in \mathbb R$ in which the constant, or the ratio between two constants, has to be in order for life to exist. 
	\item[\textbf{(ii)}] Finding the probability of such an interval, $P(I)$. 
\end{itemize}

Step (i) has been the subject of in-depth research in past decades. And, since life-permitting intervals are usually small, many say there is fine tuning. But, step (ii) --- the probability of the interval in question ---   is the sole indicator of whether or not there is fine tuning. A small $P(I)$, not a small $I$, indicates whether tuning is fine or coarse. 

Adams gives a comprehensive technical review of the state-of-the art of fine-tuning \citep{Adams2019}. His review makes clear how physics has advanced in determining what are the intervals that permit life for several fundamental constants of nature, that is step (i). But by his own reckoning, the theory still falls short of knowing how to measure the probabilities of these intervals --- step (ii). However, to say something definite about the tuning of a constant of nature (or the ratio of two such constants) $X$, with observed value $x$, whether it is either fine or coarse, the probability of its life-permitting interval $I=I_X$ must be measured. 

Step  (i) is indeed suggestive, but only after completing step (ii) a conclusion can be attained. Interval $I$ might be small and yet have a large probability. But $I$ might be very large too with very low probability. This situation, counterintuitive as it seems, has the potential of turning around some fine-tuning conclusions. Since, on the one hand, for some constant $x$, its life-permitting interval $I_X$ might be small, but $P(I_X)$ high, some constants that were regarded as finely tuned might not be. On the other hand, if $I_X$ is large, even infinitely large, but $P(I_X)$ is small, some constants, ratios of constants, or boundary conditions, formerly regarded as coarse-tuned might be fine-tuned.  McGrew and McGrew look to be the first to note that small intervals can correspond to large probabilities \citep{McGrewMcGrew2005}. Likewise, large intervals can have a small probability.

Attempts to calculate relevant probabilities have been done in the past. They all rely on a Bayesian statistical approach, whereby $X$ is regarded as a random parameter of a statistical model. The crucial issue is to determine the prior distribution of $X$. Most of the past attempts have assumed a uniform prior distribution in order to measure the relevant probability (see, among others, \citep{Barnes2017, Barnes2018, Barnes2020, Collins2012, TegmarkEtAl2006}). This is problematic because using uniformity is unwarranted when the space in which the life-permitting interval $I_X$ is taking values has infinite measure \citep{McGrewMcGrewVestrup2001, ColyvanGarfieldPriest2005}. 

%\citep{DiazHossjerMarks2021} recently proposed a theoretical framework to fine-tuning that solves these issues --- a conservative Bayesian strategy which imposes maximum uncertainty and removes all bias from the analysis.

\section{Insufficient reason and the normalization problem}

In 1933 the Russian mathematician Andrei Kolmogorov introduced the axioms of probability opening the door to rigorous mathematical research into the theory of randomness \citep{Kolmogorov2018}. 
Since these axioms are important in the discussion that follows, they are here reproduced. Let us start with a sample space $\Omega$. As its name hints, this space comprises the set from which events can be sampled. In particular, the constant of nature $X$ (or the ratio of two such constants) belongs to $\Omega$. A probability measure $P$ over $\Omega$ (that corresponds to the various values $X$ potentially could have, whether these values result in a life-permitting universe or not) must satisfy the next three axioms:

\begin{enumerate}
	\item \textbf{Unitarity:}
		\begin{align}
			P(\Omega) = 1\label{unitarity}
		\end{align}
	\item \textbf{Nonnegativitity:} For an event $E\subset\Omega$,
		\begin{align}
			P(E) \geq 0.\label{nonnegativity:}
		\end{align}
	\item \textbf{$\sigma$-additivity:}
		Any countable sequence of disjoint events $E_1, E_2, \ldots$ satisfies
		\begin{align}
			P\left(\bigcup_{i=1}^\infty E_i\right) = \sum_{i=1}^\infty P(E_i).\label{sadditivity}
		\end{align}
\end{enumerate}

For instance, when the length of $\Omega$ is finite, it is possible to assign to each interval $I \in \Omega$ a probability $\vert I \vert/ \vert \Omega \vert$, where $\vert \cdot \vert$ denotes the length.  In this way, a \textit{uniform distribution} is obtained. Of course, an uncountable number of other assignments of probabilities that satisfy Kolmogorov's axioms are also possible. However, the uniform distribution has a special place in probability. 

Jacob Bernoulli, one of the founding fathers of probability, proposed in his time what is now called \textit{the principle of insufficient reason} (PrOIR) \citep{Bernoulli1713}, also reissued by Laplace as \textit{the principle of indifference}. From the viewpoint of the continuous sample spaces of interest in cosmological fine-tuning, this principle asserts that in the absence of any prior knowledge, given that the sample space has finite length (noted as $\vert \Omega \vert < \infty$), a uniform distribution should be assumed.

The intuition behind the PrOIR has been vindicated in diverse areas like optimization and learning, Bayesian statistics, and information theory. In optimization and learning, the No-Free-Lunch theorems show that no search does better on average than a blind search, i.e., a search of the target according to uniform distributions \citep{WolpertMacReady1995, WolpertMacReady1997}. In
Bayesian statistics, the uniform probability is used as a non-informative prior \citep{Jaynes1968}; in information theory, maximum entropy provides the biggest support for the PrOIR, since the maximum entropy distribution is selected not for the negative reason of not possessing additional knowledge, but for the positive one that it minimizes the bias under the current knowledge of the problem \citep{Jaynes1957a, Jaynes1957b}. This last point is important, because, besides maximum entropy providing a positive criterium to choose the uniform distribution as the default, a generalization is allowed beyond settings in which $\Omega$ has finite length.

\subsection{Normalization}

In 2001, McGrew et al.\ offered a powerful criticism of fine-tuning arguments \citep{McGrewMcGrewVestrup2001}. They warned that even though the PrOIR can only be applied when $\Omega$ has finite length, some defenders of fine-tuning were using it to find the probabilities of interest when this length was infinite. Therefore, they argued, the PrOIR cannot be invoked in order to calculate the probabilities of most fine-tuning events, since the relevant spaces where the constants take values often have infinite cardinality. 

Their criticism is right. In the continuum, the uniform distribution is only defined for sample spaces of finite length \citep{Dembski1990}. However, in the limit uniformity ceases to be a probability distribution.\footnote{This has been known by mathematicians for  a long time. In order to motivate that a uniform distribution does not exist on the infinite-sized positive real line $\Omega= \mathbb R^+$, we first consider, $\Omega_N = (0,N]$, and assume without loss of generality that $N$ is a positive integer. Then make partition of $\Omega_N$ into $N$ subintervals of length 1:  $I_1 = (0,1], I_2=(1,2], \ldots, I_N = (N-1,N]$.  The uniform distribution assigns probability $1/N$ to each of these subintervals, since $P_N(I_i) = \vert I_i\vert/\vert \Omega_N\vert = 1/N$, for  $i \in \{1,\ldots N\}$. Therefore, as $N$ approaches infinity (noted $N \rightarrow \infty$), $\Omega_N$ approaches the whole set of positive real numbers $\Omega = \mathbb R^+$. If $P_N$ were to approach a limiting distribution $P$ on $\Omega$, then $P$ would have a continuous distribution since $P(\{x\})\le \limsup_{N\to\infty} P_N(I_{[x]}\cup I_{[x]+1}) = 0$ for any $x\in\Omega$, with $I_0$ interpreted as the empty set.  But since $P$ (if it exists) has a continuous distribution it follows that   $P_N(I_i) \rightarrow 0 = P(I_i)$, for all $i \in \mathbb N$. From this, 
\begin{align}
	1 = P(\Omega) = \sum_{i=1}^\infty P(I_i) = 0,\label{NormProb}
\end{align}	
as $N \rightarrow \infty$. That is, in the limit, $1=0$. This is a contradiction. (Notice that the first equality was obtained by unitarity \eqref{unitarity} and the second by $\sigma$-additivity \eqref{sadditivity}.) The reason for the contradiction is that the sequence of probability measures $\{P_N\}$ does not satisfy a property called {\it tightness} \citep[pp.~7-13]{Billingsley1999}, since the probability mass escapes to infinity. This implies that $\{P_N\}$ does not converge to any limiting probability measure $P$, in particular not to a uniform distribution on $\Omega$.} 

This does not mean, however, that it is impossible to obtain a probability distribution for $\Omega = \mathbb R^+$. For example,  the assignation of a distribution function $F(x)=P((0,x]) = 1-e^{-x}$ for each $x \in \mathbb R^+$ defines $P$ uniquely. For this reason we refer to $F$ as the distribution that $P$ corresponds to, or the prior distribution of $X$. The above mentioned choice of $F$ is known as the ``exponential distribution with mean parameter 1'' and it satisfies Kolmogorov's axioms.\footnote{In mathematical terms, $P$ is well defined. In fact, under the exponential distribution with mean parameter 1,
\begin{align*}
	P(I_i) = P((i-1, i]) = P((0,i]) - P((0,i-1]) = e^{-i+1} - e^{-i}.
\end{align*}
Therefore,
\begin{align*}
	1 = P(\Omega) = \sum_{i=1}^\infty P(I_i) = \sum_{i=1}^\infty \left( e^{-i+1} - e^{-i} \right) \rightarrow e^0 = 1.
\end{align*}}
Thus, it is not possible to assign a \textit{uniform} distribution to a sample space of infinite length. But, as showed, it would be mistaken to assume from this that no distribution at all exists for such a space.

Indeed, there is always the possibility to have a small length but a large probability \citep{McGrewMcGrew2005}. Nevertheless, the opposite situation is also true: it is possible for a set to have large length but small probability.\footnote{For instance, continuing with the example of the exponential distribution of mean 1, even though the interval $I = \left(10^{20}, \infty \right)$ has infinite length, $P(I) = \exp\left(-10^{20}\right) \approx 0$. On the other hand,  although $I_1 = (0,1]$, for which $\vert I_1\vert/\vert\Omega\vert = 1/\infty = 0$, the probability of $I_1$ is $P(I_1) = 1- e^{-1} \approx 0.63$.}  

This observation leads to a whole reconfiguration of the fine-tuning problem, since it is perfectly possible that some constants with small life-permitting intervals are not fine-tuned (not having small probabilities), but it is also possible that some constants whose life-permitting intervals are large, even infinite, are very finely tuned (having very small probabilities). Although in several cases a small interval leads indeed to a small probability, this is not always the case. In the end it is not size that matters, but probability.

\section{The way out - maximum entropy}\label{S:Maxent}

The appeal of the PrOIR resides in the fact that, since no better knowledge is at hand, all intervals of the same size must have the same probability. Otherwise an unwarranted addition of information would be committed \citep{DembskiMarks2009b}. However, the principle does not apply when $\Omega$ has infinite length. In fact, since no uniform distribution is possible in this scenario, some intervals of identical size will have different probabilities. Is there any way out? Is there a more general principle that supersedes Bernoulli's PrOIR? %The answer is yes --- the principle of maximum entropy.

In two subsequent papers, Edwin T. Jaynes proposed the principle of maximum entropy (maxent) which, among other things, subsumed the PrOIR  \citep{Jaynes1957a, Jaynes1957b}. There are two main differences between maxent and the PrOIR: First, maxent selects its distribution for the positive reason that it maximally reduces bias among all distributions which share the same amount/lack of information (it is ``maximally noncommittal with regard to missing information'', as Jaynes put it), instead of the negative one of selecting a distribution for not having reasons to do otherwise. Second, maxent can be generalized to distributions in unbounded spaces, provided that appropriate restrictions over some of its moments are placed  (see Appendix \ref{A:Entropy} and \citep{DiazMarks2020a, ParkBera2009}). Nevertheless, when $\Omega$ is a bounded interval, maxent still coincides with the PrOIR in that it selects the uniform distribution over $\Omega$ if no moment constraints are imposed.\footnote{In more detail, let $\Omega=[a,b]$. Among all continuous distribution $F$ with $F([a,b])=1$, the continuous version $H_c(F)$ of the entropy (defined in \eqref{Hc} of Appendix \ref{A:Entropy}) is maximized by the uniform density 
$$
f(x) = \frac{1}{b-a}, \quad a\le x \le b,
$$
when no moment restrictions are imposed. This follows by putting $d=0$ in (\ref{Lagrange}). On the other hand, when $\Omega={\mathbb R}$ is unbounded, it is necessary to impose at least one moment restriction ($d\ge 1$) on $F$ in order for $f$ to be integrable.} 

Note that informational entropy is analogous to thermodynamical entropy, but the two concepts are not equivalent. Abstract informational systems move towards maximum entropy configurations even in unbounded spaces, to the point of reaching the limit; whereas physical systems only take place in finite settings that at most can be approximated to maximum entropy distributions in unbounded spaces without ever reaching the limit.\footnote{For instance, pressure decreasing from the Earth as a function of distance can be approximated by a limiting exponential distribution, but it does not reach the limit. See Section \ref{SubS:RightS} below.} In fact, if $Y_1, Y_2, \ldots$ is a sequence of independent and identically distributed random variables with variance 1, the entropy of the normalized sum 
\begin{align}\label{NormSum}
	X = X_n = \frac{Y_1 + \cdots Y_n}{\sqrt n}
\end{align} 
is monotonically increasing \citep{ArsteinEtAl2004}. Moreover, the entropy of this sum converges to the entropy of a standard Gaussian random variable \citep{Barron1986}.\footnote{The standard Gaussian distribution maximizes entropy (is maxent) over all distributions in $\mathbb R$ with variance 1; this is a consequence of \eqref{Lagrange} in Appendix \ref{A:Entropy}, with $d=2$, $M_1(x)=x$, $\theta_1=0$, $M_2(x)=x^2$, and $\theta_2=1$.} The distribution $F_X$ of $X$ in \eqref{NormSum} can be interpreted as prior knowledge about $X$, and this distribution is the cumulative effect of $n$ components $Y_1,\ldots,Y_n$. A higher entropy of $F_X$ corresponds to less prior knowledge about $X$. Therefore, the fact that the entropy of $F_X$ increases with $n$, towards the entropy of a standard normal distribution, indicates that the prior knowledge about $X$ decreases the more random components it involves. 

In conclusion, under some restrictions, the maxent distribution over a relevant space maximizes the amount of uncertainty, or, analogously, minimizes the amount of prior knowledge. It is therefore the most conservative choice of prior distribution to make. The principle of maxent also allows the generalization of the PrOIR to unbounded spaces. These features will be exploited to solve the normalization problem.

\section{Back to fine-tuning}

To understand fine-tuning, its underlying reasoning must be grasped first. To do so, imagine a constant of nature, let us call it $x$. Fine-tuning assumes that $x=x_\text{obs}$ is a realization of a random variable $X$ and that there is an interval $I=I_X$ where $X$ must be realized for the universe to permit life. For instance, the gravitational constant $x=G_\text{obs}$ must belong to $I_G=(6.67408 \pm 0.00031) \times 10^{-11}\text m^3\cdot \text{kg}^{-1}\cdot \text s^{-2}$ to permit life \citep{XueEtAl2020}. From the perspective of fine-tuning, $x$ is a realization of a random variable $X=G$ defined in a pre-specified sample space. 

Another possibility is to have a ratio of two constants. Ratios add another layer of complexity, because tuning should be measured by considering each term in the ratio as a random variable. Gravity, for instance, is usually not considered alone in fine-tuning analysis. Rather, it typically appears in some ratio involving another constant of nature. For instance, Paul Davies considers the ratio $x=G_\text{obs}/H^2_\text{Pl}$, where $H_\text{Pl}$ is Hubble's constant at the Planck time after the big bang \citep[pp.~88--89]{Davies1982}.

With these considerations, the following procedure to measure the tuning probabilities was proposed in \citep{DiazHossjerMarks2021}:

\begin{enumerate}
	\item Determine the right sample space $\Omega$ (e.g.\ if it is discrete, continuous, finite, infinite, $\mathbb N$, $\mathbb R^+$, $\mathbb R$, $\mathbb R^n$, etc.)
	\item Determine the constraints on the distribution $F$ of $X$ (e.g., if some event $E \subset \Omega$ has a known probability; if some of its moments are finite, or finite {\it and} known; etc.).\footnote{More formally, these constraints are expressed as 				$E[M_i(X)]=\theta_i$ for $i=1,\ldots,d$. A known probability $\theta_i$ for the event $E$ corresponds to choosing $M_i(x)= \mathbf 1(x\in E)$, whereas an ordinary moment restriction corresponds to $M_i(x)=x^i$. Appendix \ref{A:Entropy} explains how the maxent distribution $F$ of $X$ is obtained from these constraints.}
	\item For $\Omega$ in Step 1 and the constraints in Step 2, find the family ${\cal F}$ of maxent distributions $F$.
	\item Find the maximum probability $\text{TP}_{\text{max}} = \max \{F(I); \, F\in {\cal F}\}$ for the life-permitting interval $I$, over the family of distributions found in Step 3.
\end{enumerate}

This four-points structure will be used in the present paper to explain concepts, but it will also be extended in many ways. The four steps are described in more detail in Subsections \ref{SubS:RightS}-\ref{SubS:ProbLPI}. For Step 1, Subsection \ref{SubS:RightS} argues extensively why an infinite sample space is warranted. For Step 2, Subsection \ref{Sec:Constraints} interprets the constraints on $F$ as prior knowledge. Subsection \ref{SubS:Maxent} explains how optimization under side constraints is used in Step 3 to find a class ${\cal F}$ of prior distributions, and proposes a solution for the lack-of-invariance problem of maxent distributions. Finally, Subsection \ref{SubS:ProbLPI} explains how Step 4, with its upper bounds on the maximal tuning probability, solves problems like the weak anthropic principle and the selection of a single point with a maxent distribution.

\subsection{Determining the right sample space}\label{SubS:RightS}

Trivial as it seems, this step is at the core of the normalization objection, since some have proposed working with sample spaces of finite length. The ad-hoc nature of a space forced to have finite length has been criticized by some authors \citep{McGrewMcGrewVestrup2001, ColyvanGarfieldPriest2005}. Barnes, on the other side, presents two arguments defending the choice of a uniform sample space  \citep{Barnes2018}. First, he subscribes to Jaynes' philosophy of not working with infinities in probability that are not, in his opinion, well-defined limits \citep[Ch.~15]{Jaynes2003}. Second, Barnes defends that the limits in which our theories can be tested before they break must also serve as limits for the sample spaces where the random variables producing the constants could take their values. Sections \ref{SubS:LimObs} and \ref{SubS:Phil} review these two objections against infinite samples spaces (or more generally, objections against working with infinities) and argue that both of them have deficiencies. Then Section \ref{SubS:Asymp} highlights that allowance of infinities opens the door to very useful asymptotic methods. 

\subsubsection{On the limits imposed by scientific observation}\label{SubS:LimObs}

Consider the second point Barnes raises. This is a common position in physics, perhaps due to its experimentalist nature. In fact, Azhar and Loeb validate Barnes' reasoning \citep{AzharLoeb2018}. Barnes cites as example the Cavendish experiment, in which, to measure gravity, a mass hangs from a rod. According to him, since the experiment does not allow for an arbitrarily large mass, because it will break the rod, larger values of the gravitational constant should not be allowed.  However, such an argument is unconvincing. To assume first that a given constant of nature $x$ is a realization of a non-degenerate random variable $X$ and then to proceed by arbitrarily limiting the values that $X$ can take might seem far-fetched.\footnote{A  random variable $X$ is degenerate if $P(X = x) = 1$ that is, if $X$ is constant with probability 1; which is a maximum entropy distribution for the restriction $M_i(y) = \mathbf 1\left(y\in \{x\}\right)$. On the other hand, $X$ is a non-degenerate random variable if it is not degenerate.} From a fine-tuning perspective, if very large values of the gravitational constant would cause the universe to crash against itself, the right conclusion is not that the gravitational constant cannot take such values, but that such hypothetical universe, not allowing for the viability of the universe, does not permit the existence of life. Therefore, the probability of $X$ attaining these values must be taken into account. 

\subsubsection{On the philosophical objections to infinities}\label{SubS:Phil}

As for Barnes' first point, Ellis et al have defended the non-existence of actual infinities in physical reality, but mention that Weierstrass' $\epsilon$-$\delta$ prescription resolved mathematical paradoxes, and thereby opened the way to a rigorous study of the field of mathematical analysis \citep{EllisEtAl2018}. We wholeheartedly agree. It is obvious that actual infinites cannot be observed in the real world, but nothing precludes its careful use in mathematics, in particular in probability. Rejecting infinites because some people have made mistakes while manipulating them is like throwing the baby out with the bath water.

Had calculus and mechanics waited until the formalization of the concept of limit (the $\epsilon$-$\delta$ prescription), they would ever have seen the light of day.  Despite the fact that calculus \emph{depends} on the concept of limit, neither Newton nor Leibniz formally defined it. Not even Euler. The definition of limit was only formalized by Cauchy more than 150 years after the invention of calculus \citep{Grabiner1983}. Science is an edifice built step by little step, and no theory is mature at birth.\footnote{In fact, no scientific theory has ever been final!} %Thus, in spite of Jaynes' many good contributions to science, like informational maximum entropy, his biases blinded him to the reality that science is an edifice built step by little step, and no theory has been final from the beginning.%\footnote{In fact, no theory has ever been final, but that is a matter for a separate discussion. However, it is related to Jaynes' strange opinion about G\"odel's incompleteness theorems, assuming that a science more oriented to information would dismiss them as a ``platitude'' \citep[p.~47]{Jaynes2003}. Jaynes' misunderstanding seems to be that new axioms could make a formal system complete, which is true, but at the cost of making the new system incomplete. Moreover, contrary to Jaynes' expectations, a better comprehension of information has actually deepened the problems that G\"odel brought to light, as shown by the famous $\Omega$ number, derived from Gregory Chaitin's algorithmic information theory \citep{Chaitin2002}. Interestingly, G\"odel's proof depends on a construction of an infinite set of theorems from a finite set of axioms, something that presumably would make Jaynes very uncomfortable.} 

Now, in probability, criticisms of treatments of infinity usually focus on the $\sigma$-additivity axiom \eqref{sadditivity}, arguing that it must be replaced by a simpler additivity axiom:
\begin{align}\label{additivity}
	P\left(\bigcup_{i=1}^n E_i\right) = \sum_{i=1}^n P(E_i),
\end{align}
for $n\in\mathbb N$; that is, additivity only over a finite number of events. This is the approach taken by the so-called Bayesian de Finetti school of probability \citep{deFinetti2008}.

However, Jaynes, a Bayesian himself, starts his book by saying: ``[W]hen all is said and done, we find ourselves, to our own surprise, in agreement with Kolmogorov and in disagreement with his critics, on nearly all technical issues... In short, we regard our system of probability as not contradicting Kolmogorov's, but rather seeking a deeper logical foundation that permits its extension in the directions that are needed for modern applications'' \citep[p.~xxi]{Jaynes2003}. What Jaynes means is that he agrees with Kolmogorov's $\sigma$-additivity axiom of probability \eqref{sadditivity}. Moreover, Jaynes even criticizes the de Finetti school for its rejection of $\sigma$-additivity and its devious use of finite additivity that leads to many paradoxes. 

In fact, Jaynes referred to William Feller, one of the main contributors to modern probability in the 20th century and a staunch critic of the de Finetti school \citep{Feller1968, Feller1971}: ``Feller saw this instantly, warned the reader against it, and proceeded to develop his own theory in a way that avoids the many useless and unnecessary paradoxes that arise from it.'' And Jaynes followed this comment by a footnote: ``Since we disagree with Feller so often on conceptual issues, we are glad to agree with him nearly in all technical ones. He was, after all, a very great contributor to the technical means for solving sampling theory problems, and practically everything he did is useful to us in our wider endeavors'' \citep[p.~466]{Jaynes2003}.

Thus, despite Jaynes' idiosyncrasies, he found himself agreeing with Kolmogorov and Feller, two of the major contributors to modern probability theory, on basically all technical aspects. Moreover, since consistency is the name of the game in mathematics, and Feller's development avoided paradoxes, there seems to be no reason to reject the modern formal measure-theoretic approach to probability.

Probability indeed developed in much more formal ways to escape many of these paradoxes. For instance, in order to avoid conditioning on events of zero probability (one of the Jaynes' biggest concerns, as dividing by zero would be equivalent to multiplying by infinity),  Doob defined conditional probability with respect to (sub)-$\sigma$-fields \citep{Doob1990}. He was following Kolmogorov's previous construction in terms of the Radon-Nikodym derivative \citep{Kolmogorov2018}. Measure theory does not disguise infinites, as Barnes suggests \citep{Barnes2018}, anymore than high-school algebra would disguise an infinite in $y = 1/x$ if care is not taken with the domain of the function. In fact, the measure-theoretic approach using the Radon-Nikodym derivative allowed to circumvent the problems of dividing by 0  (see for instance \citep[p.~343]{Resnick2014} and Appendix \ref{A:Conditional}).

\subsubsection{On the usefulness of asymptotic results}\label{SubS:Asymp}

From a practical viewpoint, both mathematically and scientifically, proper uses of infinity in probability and statistics have been extremely important. Moreover, most of these developments are continuously used in applications by physicists. All asymptotic theory is by definition sustained on the proper application of infinity, and typically the asymptotic framework requires infinite sample spaces and/or data sets of infinite size. For instance, the theory of statistical consistency says that an estimate of a given parameter is a good approximation to its true but unknown value as long as the dataset is large enough \citep{Vapnik1998}. 

To mention but one important example, Brownian motion (whose theory was pushed forward by Einstein as a physical application \citep{Einstein1905}), with increments assumed to have a normal distribution and therefore also having an unbounded sample space, is the scaling limit of a random walk ---a property known as the invariance principle. This principle has been extremely fruitful also for theoretical developments both in the finite random walk setting and for the approximating Brownian motion limit. For instance, probabilists  draw inspiration from Brownian motion intuitions for their treatment of random walks (see, e.g., \citep{Popov2021}). The invariance principle is ``a major tool in deriving results for random walks from those of Brownian motion, and vice versa. Both directions can be useful: In some cases the fact that Brownian motion is a continuous time process is an advantage over discrete time random walks\ldots In other cases it is a major advantage that (simple) random walk is a discrete object and combinatorial arguments can be the right tool to derive important features'' \citep[p. 4]{MortersPeres2010}.

Moreover, the fact that the entropy of the normalized sum \eqref{NormSum} {\it increases} towards the entropy of a standard normal distribution (which in turn is maxent over all distributions in $\mathbb R$ with mean 0 and variance 1), in agreement with the central limit theorem, strongly suggests  that the entropy of continuous distributions is a meaningful concept even for spaces of infinite Lebesgue measure.

\subsection{Determining the constraints}\label{Sec:Constraints}

Constraints are needed to find the distribution that best explains the current state of knowledge. Despite the name, these types of constraints serve a good purpose. They mean knowledge ---and since knowledge constrains randomness, knowledge also reduces uncertainty \citep{HossjerDiazRao2022}. Step 2 of the procedure for finding tuning probabilities determines which probability distributions to use. As such, different types of apriori knowledge (constraints) will lead to different selections of probability distribution functions, even over the same sample space, and such differences might affect in turn the probability of the life-permitting interval.

Now, knowledge that a restriction exists {\it and} the exact value of such a restriction sets a bar so high it is seldom reached. Of course if these two things are known, such knowledge must be used. However, in practice some kind of restriction is known to apply, while its particular value is unknown. For instance, it might be the case that $X$ has a finite expected value, $E(X)<\infty$, but there is not enough information to determine the exact value of $E(X)$. %In fact, this assumption will allow us to work in the next section under a Bayesian approach that will in turn also solve criticisms of the so-called weak anthropic principle.

More generally, the parameters of the distributions are usually some of their moments. This generates an interesting dynamic in that fine-tuning is a study of the parameters (such as $X$) of a physical model that are randomized, but that to be analyzed probabilistically one needs to assume the existence of other hyperparameters $\theta$ that govern the distribution of $X$. Then it is possible to meta-analyse the tuning of this new set of hyperparameters of the random model, going up in never-ending loops. This is in fact an inescapable epistemological situation at the very heart of mathematics and logic \citep{Carroll1895, Godel1962, Hofstadter1999}, and only a problem from the reductionist viewpoint that Jaynes seems to embrace. Nonetheless this reductionist approach will not hold water. A better approach, as has been acknowledged by Nobel Prize winners like Phillip Anderson and Robert Laughlin, is to take different levels of complexity studying problems according to each level, because ``more is different'' \citep{Anderson1972, LaughlinPines1999}. In the current setting, it means that some problems can, and must, be studied at the level of the parametric space, that is, the sample space $\Omega$ for $X$, others at the level of the space $\Theta$ of the hyperparameters, others at the level of hyper-hyper parameters $\xi\in\Xi$ that govern the distribution of the hyperparameters $\theta\in\Theta$, and so on.

Through this lens, a crucial part of the analysis of tuning is mathematical modeling, and cosmological fine-tuning is one particular instantiation. Every mathematical model, whether developed for theoretical or applied purposes, will require parameters, and those parameters can always be analyzed from a fine-tuning perspective \citep{ThorvaldsenHossjer2020, HossjerDiaz2022}.

\subsection{Find the family of maximum entropy distributions}\label{SubS:Maxent}

The election of the prior probability distribution $F$ of $X$, or the family ${\cal F}$ of prior probability distributions, to consider must be made by giving a probability of occurrence that is as fair as possible with respect to all the unknown options. Therefore, relevant knowledge should reduce the uncertainty, and this reduction must be reflected in the election of the distributions. This is attained when the distribution of maximum entropy is selected. Maximum entropy distributions keep fixed what is known ---the constraints in Step 2--- while randomizing as much as possible all that is not known. In this sense, the distribution function of interest represents the known unknowns (see, e.g., \citep{HaugMarksDembski2021}).\footnote{Take, for instance, $\Omega = \{x_1,x_2,x_3\}$. In the absence of further knowledge, the Shannon entropy $H(F)$ in (\ref{HF}) in Appendix \ref{A:Entropy} is maximized by a uniform distribution $F$ on $\Omega$, with $\pi_1=\pi_2 =\pi_3= 1/3$, and $\pi_i = F(\{x_i\})$. However, let us assume that $\pi_1= E[M_1(X)]=1/2=\theta_1$, with $M_1(x)= \mathbf 1(x=x_1)$, represents information that is known to the researcher. Under this constraint $H(F)$ is maximized by $(\pi_1,\pi_2,\pi_3)=(1/2,1/4,1/4)$. Therefore, the knowledge of the probability of the event $\{x_1\}$ fixes the uncertainty for this event, and forces the redistribution of the remaining probability as fair as possible among the remaining two options $x_2$ and $x_3$.}

There are three possibilities here:

\begin{enumerate}
	\item A distribution function of maximum entropy does not exist ($\cal F$ is empty).
	\item The knowledge we possess (the restrictions we add) produces only one distribution function (${\cal F}=\{F\}$ contains one single distribution $F$).
	\item The knowledge we possess admits more than one distribution function ($\vert \cal F\vert$ $> 1$).
\end{enumerate}

In the first case, the constraints are contradictory so that a maximum entropy distribution does not exist. Imagine, for instance, that the researcher requires a negative expected value for a random variable that is only positive, or a positive variance of a degenerate random variable. This means that the assumptions, or the model itself, needs to be reformulated. However, since the constant of nature does exist, at the very least the degenerate random variable must exist.\footnote{In this scenario, there is no fine-tuning. For a constant of nature with value $x$, this corresponds to choosing $d=1$ and $M_1(y) = \mathbf 1\left(y \in \{x\}\right)$, which in this scenario corresponds to $\theta_1 = E[M_1(X)] = P(X=\{x\}) = 1$ (see footnotes 6 and 7 and \eqref{Lagrange} in Appendix \ref{A:Entropy}).} Nevertheless, by the very essence of the fine-tuning problem, the observed constant of nature $x=x_{\text{obs}}$ is  assumed to be the realization of a random variable $X$ that can take multiple values. Therefore fine-tuning requires a non-degenerate random variable, which in turn implies a non-zero scale parameter (e.g., a positive variance, when it exists).

The second case ensues when knowledge warrants the selection of a single maximum entropy distribution. This is the case, for instance, when the constraints of Step 2 correspond to $d$ moments of $F$, and these are all known. It occurs when there is a single ($d=1$) constraint on the expected value of $X$ and the sample space is the nonnegative portion of the real line and additionally the mean is known to be $\theta$. Then the unique maximum entropy distribution over this space corresponds to the exponential distribution with mean parameter $\theta$.\footnote{This corresponds to choosing $d=1$ and $M_1(x)=x$ in \eqref{EMi}.} Or if $X$ is a Weibull random variable with scale parameter $\theta_1 = \lambda>0$ and shape parameter $\theta_2= k>0$, it has maximum entropy distribution over all distributions in $\mathbb R^+$ such that $E(X^k) = \lambda^k$ and $E(\ln X) = \ln \lambda - \lambda_E/k$, where $\lambda_E$ is the Euler-Mascheroni constant:\footnote{This corresponds to choosing $d=2$, $M_1(x)=x^k$ and $M_2(x)=\ln (x)$ in \eqref{EMi}.}
\begin{equation*}
	\lambda_E = \lim_{n \rightarrow \infty} \left( -\ln n + \sum_{i=1}^n\frac{1}{i}  \right).
\end{equation*}
In a similar way, different maximum entropy distributions are produced under various restrictions over the same space  \citep{ParkBera2009}.

As for the third case, when there is knowledge that a restriction exists but the exact value of the restriction is not known, a family of maximum entropy distributions is obtained. Continuing with the previous example, let $\Omega = \mathbb R^+$. If the researcher only knows that the random variable $X$ has finite first moment $E(X)=\theta$, but the exact value of $\theta$ is unknown (i.e., $E(X)<\infty$), then there is a family of maximum entropy distributions to consider, not only a single one. In this particular case, an uncountable number of exponential distributions with densities $f(x;\theta)=e^{-x/\theta}/\theta$ results, where $\theta$ takes values along the positive real line, i.e.\  $\Theta=\mathbb R^+$.

More generally, for $\theta = (\theta_1,\ldots, \theta_d)$, a finite-dimensional vector of hyperparameters in a given parameter space $\Theta\subset \mathbb R^d$, the maximum entropy distribution of $X$ belongs to a class 
\begin{equation}\label{class}
	\mathcal F = \{F\left(\cdot\, ;\, \theta\right), \text{ such that } \theta\in\Theta \}, 
\end{equation}
where $F(\cdot\, ; \theta)$ is the maxent distribution corresponding to the $d$ moment constraints 
\begin{equation}
    E[M_i(X)] = \theta_i
    \label{EMi},
\end{equation}
for $i=1,\ldots,d$. For instance, by an appropriate choice of $d$, $M_i$ and $\theta_i$, this provides a solution for fine-tuning to the problem of maxent not being invariant with respect to transformations of data \citep{Koperski2005}.\footnote{In more detail, suppose $Y_i=G_i(X)$ is a strictly increasing and differentiable transformations of $X$, with $g_i(x)=G_i^\prime(x)$, for $i=1,\ldots,d$. Then, if $F$ and $F_i$ refer to the distributions of $X$ and $Y_i$, and $f=F^\prime$ is the density of $X$, it can be shown that 
\begin{align}\label{HcFi}
H_c(F_i) = H_c(F) + E[M_i(X)],
\end{align}
where $M_i(x)=\log[g_i(x)]$, and $H_c$ is the continuous entropy of \eqref{Hc} in Appendix \ref{A:Entropy}. The maxent distribution $F_i$ therefore corresponds to the distribution $F$ that maximizes $H_c(F)+\theta_i$ when $\theta_i=E[M_i(X)]$ varies. From this it follows that the maxent distributions of $F_1,\ldots,F_d$ are all members of ${\cal F}=\{F(\cdot;\theta); \,\,\theta =(\theta_1,\ldots,\theta_d)\in \Theta\}$, with $\Theta$ ranging over all permissible constraints on $(E[M_1(X)],\ldots,E[M_d(X)])$.}

\subsection{Find the probability of the life-permitting interval}\label{SubS:ProbLPI}

Remember that the goal is to calculate the probability of a constant of nature $X$ falling in its life-permitting interval $I_X$. To do this, with a Bayesian approach, the observed value $x_\text{obs}$ of this constant is taken as a realization of a random variable $X$, and the event of interest is $\{X\in I_X\}$. Having found the relevant family ${\cal F}$ of prior distributions of $X$ in Step 3, the tuning probabilities
\begin{equation}
	\mbox{TP}(\theta) = F(I_X;\theta)
\end{equation}
that $X$ is tuned to lie in $I_X$, for different $\theta$ in $\Theta$, are obtainable. The conservative approach is then to take the value $\theta_\text{max}$ that maximizes the tuning probability $\mbox{TP}(\theta)$, and to put
\begin{equation}\label{TPmax}
	\mbox{TP}_\text{max} = \max_{\theta\in\Theta} \mbox{TP}(\theta) = F\left(I_X; \theta_\text{max} \right).
\end{equation}

Note in particular that the maximal tuning probability \eqref{TPmax} does not depend on any distributional assumptions on the hyperparameters $\theta$ in terms of hyper-hyperparameters $\psi$. Thereby $\mbox{TP}_\text{max}$ to some extent avoids the loop of hierarchical meta-assumptions, described in Section \ref{Sec:Constraints}. However, the parameter space $\Theta$ must still be chosen apriori, with which the hierarchy under analysis is pre-specified. 

When $d=1$ and $\theta=E(X)$, taking the observed value $x_{\text{obs}}$ of the constant of nature as estimator $\hat{\theta}=x_\text{obs}$ of the mean of the random variable $X$ to find an estimate $\hat{F}(\cdot)=F(\cdot;\hat{\theta})$ of the maxent distribution $F$ over $\Omega$ with mean $x_{\text{obs}}$ might be tempting. That is, using a sample of size 1. There is nothing intrinsically wrong in statistical terms with samples of size 1. In fact, a sample of size 1 from $F$ is still an unbiased estimator of the first moment of $F$. Now, according to the {\it weak anthropic principle}, this single observation is biased in a deeper sense: since we live in a habitable universe, we are constrained in our sample of size 1 to only observe values that permit life. The weak anthropic principle is thus a problem of selection bias \citep{Bostrom2002}. The setting in \eqref{TPmax} provides the full resolution to the weak anthropic principle for any number $d$ of moment constraints. Namely, instead of taking the observed value of the constant of nature in our universe to determine the probability, all the possible maxent distributions in the family $\mathcal F$ are taken into account, and the one that is selected maximizes the probability of $\{X \in I_X\}$. %Therefore, concerns related to the weak anthropic principle are also dismissed. 
In fact, when $d=1$ and $\theta=E(X)$, \eqref{TPmax} also addresses and solves another concern by McGrew, according to which a unique maxent distribution is assumed and then a single observation $x_{obs}$ of $X$ is taken as a possibly biased estimate of the mean $E(X)$ of $X$ \citep{McGrew2018}. %Our approach does not take a single distribution but a family of maxent distributions, and then, conservatively, selects that which gives the interval $I_X$ the highest probability.

In proper terms then, \eqref{TPmax} does not obtain an exact tuning probability, but an upper bound to it. This, among other things, implies that whenever this method detects fine-tuning for a particular model (i.e., a low probability of $X \in I_X$), fine-tuning is indeed present. However, when it detects coarse-tuning (i.e., a high probability of $X\in I_X$), there is no guarantee that the actual tuning is coarse. The explanation is that, when taking the conservative approach of selecting the value $\theta_\text{max}$ that maximizes the probability of the event $\{X \in I_X\}$ over the family of distributions ${\cal F} = \{F\left(\cdot\, ;\, \theta\right);\, \theta\in\Theta\}$, then 
\begin{equation*}
	\mbox{TP}_\text{max} \ge \mbox{TP}(\theta_\text{real}),
\end{equation*}
where $\theta_\text{real}$ corresponds to the actual value of the parameter. Therefore, if $\mbox{TP}_\text{max}$ is small, $\mbox{TP}(\theta_\text{real})$ is also small. However, if $\mbox{TP}_\text{max}$ is large, the procedure cannot detect whether $\mbox{TP}(\theta_\text{real})$ is large or small. 

To put it in statistical terms, if negatives and positives correspond to fine and coarse tuning respectively, the rate of false negatives is zero whereas the rate of false positives cannot be determined. Thus, when $\mbox{TP}_\text{max}$ is large, it does not mean that there is coarse-tuning, but that the four-steps method of Section \ref{SubS:Maxent} is inconclusive for determining fine-tuning.

\section{Examples}

In this section the life permitting interval of $X$ is assumed to have the form
\begin{equation}
    I = I_X = x_\text{obs} [1-\epsilon,1+\epsilon],
    \label{IX}
\end{equation} 
when the observed value $x_\text{obs}$ of $X$ is nonzero, with $\epsilon>0$ a dimensionless constant that quantifies half the relative size of $I_X$. Appendix \ref{A:UpperBounds} gives general formulas for the maximal tuning probability 
\begin{equation}
\mbox{TP}_\text{max} = \mbox{TP}_\text{max}(\epsilon,{\cal F})
\label{TPmax1}
\end{equation}
as a function of $\epsilon$ and the family ${\cal F}$ of prior distributions of $X$. 

Appendix \ref{A:UpperBounds} demonstrates that the maximal tuning probability in \eqref{TPmax1} is proportional to $\epsilon$  for one-dimensional ($d=1$) families of priors, with constants of proportionality $C_1$ for the scale-family and $C_2$ for the location-family of priors respectively; see \eqref{A:Constants}. Consequently, for these scenarios a small $\epsilon$ implies fine-tuning. 

For the two-dimensional ($d=2$) form-scale and location-scale families, the maximal tuning probability is proportional to $\epsilon\sqrt{S}$, where $0<S\le \infty$ is the maximum signal-to-noise ratio of the family of priors. The larger $S$ is, the less we assume about the family of priors, with $S=\infty$ corresponding to a scenario in which degenerate one-point prior distributions are allowed. Appendix \ref{A:UpperBounds} shows that the maximal tuning probability is proportional to $\epsilon \sqrt{S}$ as long as this product remains small, whereas it equals 1 for $S=\infty$. Consequently, an interval with a small relative size $\epsilon$ implies a small maximal fine-tuning probability if the maximum signal-to-noise ratio $S$ is of smaller order than $\epsilon^{-2}$, but possibly a large coarse-tuning probability if the order of $S$ is larger than $\epsilon^{-2}$.
 
In Sections \ref{SubS:Crit}-\ref{SubS:Prim}, three examples illustrate this approach to find the maximal tuning probability; the critical density of the universe, gravitation, and the amplitude of primordial fluctuations.   

\subsection{Critical density of the universe}\label{SubS:Crit}

According to \citep[p.~89]{Davies1982}, the critical density of the universe $\rho_\text{crit}$ cannot take values outside the life-permitting interval 
\begin{align*}
	I_{\rho_\text{crit}} = \left[\rho_\text{crit} -  \rho_\text{crit} 10^{-60}, \rho_\text{crit} +  \rho_\text{crit} 10^{-60}\right].
\end{align*} 
That is, in accordance with (\ref{IX}),  $I_{\rho_\text{crit}}$ can be expressed as $\rho_\text{crit}[1-\epsilon, 1+ \epsilon]$, for dimensionless $\epsilon$ that is half the relative size of $I_{\rho_\text{crit}}$. Since $\epsilon =10^{-60}$, it is clearly very small ($\epsilon \ll 1$). Also, since the density cannot be negative, tuning of the critical density is evaluated for $\mathbb R^+$ in Table \ref{T:Tuning}. In this case, in two of the three relevant scenarios of Table \ref{T:Tuning} the tuning is fine:

\begin{enumerate}
	\item When the distribution function of the critical density belongs to the scale family, with $\text{TP}_\text{max} = 2\times 10^{-60} C_1$, with $C_1=e^{-1}$ for the family of exponential distributions.
	\item When the distribution function of the critical density belongs to a form-and-scale family (for instance, the gamma or Weibull families), with $\text{TP}_\text{max} = 2\times 10^{-60} \sqrt{S/2\pi}$, provided the signal-to-noise ratio of the prior is bounded above by $S$ and 		$\sqrt S \ll 10^{60}$.
\end{enumerate}

As per the second row of Table \ref{T:Tuning}, fine-tuning cannot be detected when the prior distribution of the critical density belongs to the form-and-scale family and no restrictions are imposed on the signal-to-noise ratio (i.e., $S=\infty$).

\subsection{Gravitational force}\label{SubS:Grav}

When observing the ratio $X$ of the gravitational constant $G_\text{obs}$ to the contribution from vacuum energy to the cosmological constant $\Lambda_\text{vac}$, according to Davies \citep[p.~107]{Davies1982}, gravitation cannot fall outside the life-permitting interval 
\begin{align*}
	I_X = x_\text{obs}\left[1-10^{-100},1+10^{-100}\right]. 
\end{align*}

For simplicity we will refer to the ratio $X=G/\Lambda_\text{vac}$ as a gravitational random variable. Analogously  to (\ref{IX}) and the previous example, its life-permitting interval $I_X$ can be expressed as $x_\text{obs}[1-\epsilon, 1+ \epsilon]$, for dimensionless $\epsilon = 10^{-100}$, which is very small. Now, from the perspective of tuning, the gravitational constant is a more complex case than the critical density, since depending on assumptions it can be positive, negative, or even zero. Under the assumption that gravity is only an attractive force, it can only be positive, so its sample space is $\mathbb R^+$. In this scenario, analogous conclusions to those in the previous example are attained (just being careful to substitute $\epsilon$ for $10^{-100}$, instead of $10^{-60}$).

However, under the assumption that the gravitational constant could also be repulsive, it could take negative values too \citep[p.~535]{Barnes2012}. In this scenario, the sample space is the whole real line $\mathbb R$. According to Table \ref{T:Tuning}, five scenarios are possible here; in three of them fine-tuning is detected, and in two of them the criterium is inconclusive assigning a coarse $\text{TP}_\text{max} = 1$.  The three cases in which there is fine-tuning are the following:

\begin{enumerate}
	\item When the distribution of the gravitational random variable belongs to the scale family, since $0 \notin I_X$. In this case, $\text{TP}_\text{max} = 2\times 10^{-100} C_1$, with $C_1=0.5e^{-1}$ for the family of Laplace distributions and $C_1=e^{-1/2}/\sqrt{8\pi}$ for the family of symmetric normal distributions.
	\item When the distribution of the gravitational random variable belongs to the location family, in which case $\text{TP}_\text{max} = 2\times 10^{-100} C_3$, for $C_3 \ll 10^{100}$. For instance, $C_3=1/(\sqrt{2\pi}\sigma)$ for the family of normal distributions with fixed standard deviation $\sigma$. 
	\item When the distribution of the gravitational random variable belongs to the location-and-scale family,  in which case $\text{TP}_\text{max} = 2\times 10^{-100}(C_3\sqrt S + C_1)$, where $C_1 = e^{-1/2}/\sqrt{8\pi}$ and $C_3=1/{\sqrt{2\pi}}$ for the family of normal distributions. Consequently, there is fine-tuning provided the signal-to-noise ratio is bounded above by some $S$, and $S \ll 10^{200}$.
\end{enumerate}

For the cases in which fine-tuning is not detected the distribution of the gravitational random variable must belong either to the location or the location-and-scale families of distributions with small enough $\sigma$ and large enough $S$ respectively. The scenario in the fifth row of Table \ref{T:Tuning} does not apply in this situation, since $0 \notin I_X$. 

\subsection{Amplitude of primordial fluctuations}\label{SubS:Prim}

The amplitude of the primordial fluctuations must be in a life-permitting interval $I_Q = \left[10^{-6}, 10^{-5}\right]$ \citep[p.~128]{Rees2000}. However, this interval spans over more than one order of magnitude and for this reason the theory from Appendix \ref{A:UpperBounds} for small $\epsilon$ does not apply (see Remark \ref{A:epsilon} on Appendix \ref{A:Summary}). In fact, under the assumption that the amplitude of the primordial fluctuations has an exponential distribution (corresponding to the scale family of distributions whose sample space is $\mathbb R^+$),  $\text{TP}_\text{max} \approx 0.7$ \citep{DiazHossjerMarks2021}. Therefore, surprisingly, fine-tuning is not detected in this case.

\section{Discussion}

This article discusses a method that formalizes the detection of fine-tuning and solves many problems related to the measurement of fine-tuning for the constants of physics. Namely, the normalization problem, the weak anthropic principle, the invariance of maximum entropy, and the selection of a single maxent distribution.

One of the great advantages of the maxent approach is its great versatility. Even if the constants change, or even under a different model altogether, this approach will still work. As explained in Sections \ref{SubS:RightS}-\ref{SubS:ProbLPI}, this flexibility of the maxent approach follows since the class of prior distributions of the relevant constant of nature is obtained once the constant's sample space and constraints have been defined. This is highly relevant at a time when the standard model of cosmology is under scrutiny (see, e.g., \citep{SecrestEtAl2021, Sarkar2022}). 

Another advantage is that fine-tuning has been attached to mathematical modeling, and cosmological fine-tuning has become just a particular application. Fine-tuning can then be used in any area of science where a mathematical model is present. For the particular case of cosmology, our framework entails that even if the current model that describes the universe is replaced by another, the method herein will still apply to the new model with its new set of constants of nature and their associated sample spaces and constraints. Thus, as long as mathematical modeling is present, particularly a mathematical model to explain how the universe works, fine-tuning is going to be a relevant question to ask. The approach taken in this paper is well-suited for studying a large class of models. 

\section*{Declarations}

Data sharing not applicable to this article as no datasets were generated or analyzed during the current study.

\appendix

\section{Entropy and maximum entropy distributions}\label{A:Entropy}

Suppose $X$ is a discrete random variable defined on a countable sample space $\Omega = \{x_1,x_2,\ldots\}$ with distribution $F$. Then the entropy of $F$ is 
\begin{align}\label{HF}
H(F) = -\sum_i \ln(\pi_i)\pi_i, 
\end{align}
with $\pi_i=F(\{x_i\})$. 

For a continuous random variable $X$, defined on a subset $\Omega$ of $\mathbb R^+$, with density function $f=F^\prime$, the entropy $H(F)$ is not defined. However, for each $\delta>0$, the distribution $F$ of $X$ can be approximated by a discrete distribution $F_\delta$ such that $F_\delta(\{x_i\}) = F(I_i)$ for $i=1,2,\ldots$, where $I_i=((i-1)\delta,i\delta]$ and $x_i$ is the mid point of $I_i$. If $\delta > 0$ is small, we have approximately 
$$
H(F_\delta) \approx  -\int_\Omega \ln[f(x)]f(x)dx + \ln [\delta^{-1}] = H_c(F) + \ln [\delta^{-1}].
$$
Although $H(F_\delta) \to \infty$ as $\delta\to 0$, we may use 
\begin{align}\label{Hc}
H_c(F) = \lim_{\delta\to 0} \left[H(F_\delta) - \ln\left(\delta^{-1}\right)\right] =  -\int_\Omega \ln[f(x)]f(x)dx
\end{align}
as a continuous analogue of the entropy. The motivation of $H_c$ as a continuous analogue of the entropy is similar for other unbounded sample spaces $\Omega$, such as $\mathbb R$, $\mathbb R^+\times\mathbb R$, and $\mathbb R^2$.  

The goal is to find a distribution $F$ that maximizes $H_c(F)$ subject to a constraint $\int_\Omega f(x)dx = 1$ and the $d$ additional moment constraints $E[M_i(X)]=\int_\Omega M_i(x) f(x)dx=\theta_i$ for $i=1,\ldots,d$. In particular, ordinary moment constraints correspond to choosing $M_i(x)=x^i$ as polynomials of $x$ of various orders $i$. The distribution $F$ with maximal entropy $H_c$, subject to these $n$ constraints, has a density function
\begin{align}\label{Lagrange}
f(x) = \frac{\exp(\lambda_1 M_1(x)+\ldots + \lambda_d M_d(x))}{Z(\lambda_1,\ldots,\lambda_d)},
\end{align}
where $\lambda_1,\ldots,\lambda_d$ are Lagrange multipliers chosen to satisfy the moment constraints, whereas $Z$ is a normalizing partition function, chosen so that $f$ integrates to 1.

Though this is the most common way to find maximum entropy distributions, a more general approach is possible that finds maximum entropy distributions, even in some cases for which Lagrange multipliers do not work \citep{Conrad2005}. The details go beyond the scope of this paper.

\section{Conditional probability}\label{A:Conditional}

In the context of tuning, suppose we have two constants of nature, $X,Y\in \mathbb R^+$. We want to find the conditional distribution $x\to F_{X \mid Y}(x) = P(X\le x \mid Y=y)$ of $X$ given an observed value $y$ of $Y$. This can be formulated as a conditional probability $P(A \mid G) = P(A \cap G)/P(G)$, where $A$ and $G$ are the sets of outcomes for which $X\le x$ and $Y=y$ respectively. However, this formula does not work when $Y$ has a continuous distribution and $P(G)=0$. Given this limitation of the classical notion of conditional probability, a more general definition of conditional probability is needed in order to find the conditional distribution of $X$ given $Y$. Given this, ``the whole point of [the measure-theoretic treatment of conditional probability] is the systematic development of a notion of conditional probability that covers conditioning with respect to events of probability 0. This is accomplished by conditioning with respect to \textit{collections} of events ---that is, with respect to $\sigma$-fields $\mathcal G$'' \citep{Billingsley1995}. %\citep[pp.~432]{Billingsley1995}. 

The formal definition of conditional probability is as follows: Given a probability space $(\Sigma,{\cal H},P)$, a set $A\in{\cal H}$, and and a $\sigma$-field $\mathcal G \subset \mathcal H$, there exists a function $f$ (whose existence is guaranteed by the Radon-Nikodym theorem), $\mathcal G$-measurable and integrable with respect to $P$, such that $P(A \cap G) = \nu (G) = \int_G fdP$ for all $G \in \mathcal G$. This function $f$ can be conveniently noted as $P(A \mid \mathcal G)$. The function $P(A \mid \mathcal G)$ thus has two properties that define it:
\begin{itemize}
	\item[(i)] $P(A \mid \mathcal G)$ is $\mathcal G$-measurable and integrable.
	\item[(ii)] $P(A \mid \mathcal G)$ satisfies the functional equation
	\begin{align*}
		\int_G P(A \mid \mathcal G) dP = P(A \cap G), \ \ \ \ \ \ \ G \in \mathcal G.
	\end{align*}
\end{itemize}
The important point here is that writing $f$ as $P(A \mid \mathcal G)$ is just a notational convenience that can be interpreted as follows: when $\mathcal G$ is generated by a partition $\mathcal P = \{G_1, G_2 \ldots\}$ of $\Sigma$, conditioning on $\mathcal G$ can thus be seen as performing an experiment whose outcome 
$$
f(x) = P(A \mid \mathcal G)(x) = \sum_{i=1}^\infty P(A \mid G_i) 1(x\in G_i)
$$
will determine which event $G_i$ of the partition occurred.\footnote{In our motivating example with two constants of nature $X$ and $Y$, a partition $\cal P$ corresponds to a case when the sample space of $Y$ is countable ($Y\in\{y_1,y_2,\ldots\}$) and $G_i$ is the set of outcomes for which $Y=y_i$.} In general, there is a whole family of random variables satisfying properties (i) and (ii). Such random variables are equal with probability 1 (i.e., they can only be different in a set of zero probability); for this reason, each member of the family is called  a \textit{version} of the others. Thus, $P(A \mid \mathcal G)$ stands for any member of this family. Therefore, if $P(G_i) = 0$ for some nonempty $G_i \in \mathcal P$, a constant value $c$ must be selected to make $P(A \mid \mathcal G) = c$ on $G_i$. Regardless of the choice of $c\in[0,1]$, $P(A \mid \mathcal G)$ can still be considered a probability measure on $(\Sigma,{\cal H})$ that assigns probabilities to all $A\in{\cal H}$. That is, a version of the conditional probability is selected for any value of $c$.

However, this interpretation for the notation $P(A \mid \mathcal G)$ does not hold when $\mathcal G$ is not generated by a partition of $\Sigma$, which in our motivating example corresponds to $Y$ having a continuous distribution. Nonetheless, even though the \textit{intuition} of the conditional probability as the realization of an experiment, with an outcome in $\mathcal P = \{G_1,G_2,\ldots\}$, is gone, the mathematical formalism stands, independently of the conditioning $\sigma$-field. That is, there still exists a family of functions $\{f\}$ satisfying properties (i) and (ii). As in any area of mathematics, problems would arise when dividing by zero, but the formalism permits to circumvent this situation by selecting a well-defined version of the conditional probability.

\section{Upper bounds for fine-tuning probabilities}\label{A:UpperBounds}

Let $x_{\text{obs}} \ne 0$ be the observed value of a constant of nature $X\in \Omega$ (or a ratio $X$ of two constants of nature), where  $\Omega$ is the sample space. The prior distribution of $X$ has a density $f=F^\prime$ that belongs a parametric family 
\begin{equation}\label{cPDefGen}
	\mathcal F = \{f(x;\theta); \,\, \theta = (\theta_1,\ldots,\theta_d)\in \Theta\}, 
\end{equation}
with $\Theta \subset \mathbb R^d$ the parameter space to which the hyperparameter $\theta$ belongs. This framework is more general than \eqref{class}, since \eqref{cPDefGen} does not assume that the hyperparameters $\theta_i$ represent moment constraints $\theta_i=E[M_i(X)]$ of Appendix \ref{A:Entropy}. Let
\begin{align}\label{IXApp}
	I = I_X = x_{\text{obs}}  [1-\epsilon,1-\epsilon] 
\end{align}
be the life-permitting interval of $X$, with $\epsilon>0$ a small number that quantifies half the relative size of $I$. For a fixed $\theta$ the tuning probability is
\begin{equation}\label{PmaxImusi}
	\text{TP}(\theta) = F(I;\theta) = \int_I f(x;\theta) dx.
\end{equation}
Our quantity of interest is a useful approximation of the upper bound
\begin{equation}\label{PmaxIGen}
	\text{TP}_{\text{max}} (I) = \max_{\theta\in\Theta} F(I;\theta)
\end{equation}
of the tuning probability, assuming that the maximum in \eqref{PmaxIGen} is taken over all hyperparameter vectors that appear in \eqref{cPDefGen}.  

Appendix \ref{AS:SPR+}-\ref{AS:LSR} will give explicit approximations of the maximal tuning probability (\ref{PmaxIGen}) for give different families ${\cal F}$ of prior distributions. Then in Table \ref{T:Tuning} of Appendix \ref{A:Summary} the results are summarized. 

\subsection{Scale parameter for \texorpdfstring{$\Omega = \mathbb R^+$}{OR+}}\label{AS:SPR+}

A scale-parameter corresponds to $d=1$ and $\theta=\sigma >0$, so that $\Theta = \mathbb R^+$. The prior density is
\begin{equation}
	f(x;\sigma) = \frac{1}{\sigma} g\left(\frac{x}{\sigma}\right).
\label{pScale}
\end{equation}

A typical example is the family of exponential distributions ($g(x)=e^{-x}$). Since 
\begin{align}\label{A:FBound1}
	F(I;\sigma) \approx 2\epsilon \frac{x_\text{obs}}{\sigma} g\left(  \frac{x_\text{obs}}{\sigma} \right),
\end{align}
it follows that 
\begin{align}\label{PmaxScale}
\begin{aligned}
		\text{TP}_{\text{max}}(I) &\le  2\epsilon \max_{x>0}\{xf(x)\}\\
		& =  2\epsilon C_1.
\end{aligned}
\end{align}
It is easily seen that $C_1=e^{-1}$ for the family of exponential distributions. 

\subsection{Form and scale parameter for \texorpdfstring{$\Omega = \mathbb R^+$}{OR}} 

This scenario corresponds to $d=2$, $\theta=(\psi,\sigma)$, where $\psi>0$ is the form parameter and $\sigma>0$ the scale parameter. Consequently 
\begin{equation}
	f(x;\psi,\sigma) = \frac{1}{\sigma} g\left(\frac{x}{\sigma};\psi\right),
	\label{pFormScale}
\end{equation} 
where $g(\cdot;\psi)$ is the density of the distribution with shape parameter $\psi$ and scale parameter 1. Typical examples are the family of gamma distributions ($g(x;\psi)=x^{\psi-1}e^{-x}/\Gamma(\psi)$) or the family of Weibull distributions ($g(x;\psi)=\psi x^{\psi -1}e^{-x^\psi}$). The expected value and variance of $X$ are
\begin{align*}
	E(X) &= \mu = h_1(\psi)\sigma,\\
	\text{Var}(X) &= h_2(\psi)\sigma^2,
\end{align*}
respectively, for some functions $h_1$ and $h_2$ (for instance $h_1(\psi)=h_2(\psi)=\psi$ for the family of gamma distributions, whereas $h_1(\psi)=\Gamma(1+1/\psi)$ and $h_2(\psi)=\Gamma(1+2/\psi)-\Gamma^2(1+1/\psi)$ for the family of Weibull distributions). Let $h(\psi)=h_1^2(\psi)/h_2(\psi)$. We consider a parameter space
\begin{equation}
	\Theta = \left\{(\psi,\sigma): \frac{E^2(X)}{\text{Var}(X)} = h(\psi) \le S\right\}
	\label{ThetaLocForm}
\end{equation}
of hyperparameters such that the signal-to-noise ratio 
\begin{equation}
	\text{SNR} = \frac{E^2(X)}{\text{Var}(X)} \le S
\label{SNR0}
\end{equation}
of the prior distribution is at most $S$.  We will prove that
\begin{equation}
	\text{TP}_{\text{max}} (I) \le 2\epsilon C_2\sqrt{S}
\label{PmaxFormScale}
\end{equation}
for some constant $C_2$ (defined below), whenever $S$ is large and
\begin{equation}
	\epsilon \sqrt{S} \ll 1 \Longleftrightarrow S \ll \frac{1}{\epsilon^2}.
	\label{PmaxCond}
\end{equation}
It follows that essentially, \eqref{PmaxCond} is the condition for the upper bound $\text{TP}_{\text{max}} (I)$ of the tuning probability to be small.  \par\medskip
{\bf Proof of \eqref{PmaxFormScale}.} Equation \eqref{pFormScale} implies
\begin{align*}
	F(I;\psi,\sigma) = F\left(\frac{I}{x_\text{obs}};\psi,\frac{\sigma}{x_\text{obs}} \right),
\end{align*}
from which 
\begin{equation}
	\text{TP}_{\text{max}} (I) = \text{TP}_{\text{max}} \left(\frac{I}{x_\text{obs}}\right)
	\label{PmaxIInv}
\end{equation}
follows. Since $I/x_\text{obs} = [1-\epsilon,1+\epsilon]$, we assume, without loss of generality, that $x_\text{obs}=1$. With this choice of $x_\text{obs}$, 
\begin{equation}
	F(I;\psi,\sigma) \approx  \frac{2\epsilon}{\sigma} g\left(\frac{1}{\sigma};\psi\right).
\label{PApprox}
\end{equation}
In view of \eqref{ThetaLocForm} and \eqref{PApprox}, 
\begin{align*}
	\text{TP}_{\text{max}}(I) &= \max_{\substack{\psi:h(\psi)\le S;\\ 
								\sigma >0}}   
						F(I;\psi,\sigma)\\
	&\approx  2\epsilon \left(\max_{\substack{\psi:h(\psi)\le S;\\ 
								\sigma >0}} 
						\left\{\frac{1}{\sigma} g\left(\frac{1}{\sigma};\psi\right)\right\}\right)\\
	&=  2\epsilon \left(\max_{\substack{\psi:h(\psi)\le S;\\ 
								x >0}} 
						\left\{x g(x;\psi)\right\}\right)\\
	&\le 2\epsilon C_2 \sqrt{S}.
\end{align*}
Moreover, it can be seen that
\begin{align}\label{CForm}
\begin{aligned}
		C_2 &= \frac{1}{\sqrt{S}}\left(\max_{\substack{\psi:h(\psi)\le S;\\ 
								x >0}} 
						\{x g(x;\psi)\}\right)\\
		&\approx \frac{1}{\sqrt{S}} \cdot  h_1\left(h^{-1}(S)\right) g\left(h_1\left(h^{-1}(S)\right);h^{-1}(S)\right)
\end{aligned}
\end{align}
when $S$ is large, since the maximum of $x g(x;\psi)$ with respect to $x$ approximately equals $E(X)g(E(X);\psi)$. For the gamma and Weibull families, we have that $g(x;\psi)$ is approximately a Gaussian density for large signal-to-noise ratios $h(\psi)$. This implies 
\begin{equation}
	g(x;\psi) \approx \frac{1}{\sqrt{h_2(\psi)}}\cdot \phi\left(\frac{x-h_1(\psi)}{\sqrt{h_2(\psi)}}\right)
	\label{gNorm}
\end{equation}
when $h(\psi)$ is large, where $\phi(x)=e^{-x^2/2}/\sqrt{2\pi}$ is the density of a standard normal distribution. Inserting \eqref{gNorm} into \eqref{CForm} we find (with $\psi_1 = h^{-1}(S)$) that 
\begin{align*}
	C_2 \approx \frac{1}{\sqrt{S}}\cdot  h_1(\psi_1) \cdot  \frac{1}{\sqrt{h_2(\psi_1)}}  \phi\left(\frac{h_1(\psi_1)-h_1(\psi_1)}{\sqrt{h_2(\psi_1)}}\right) = \frac{1}{\sqrt{2\pi}},
\end{align*}
where in the last step we used that $h_1(\psi_1)/\sqrt{h_2(\psi_1)}=\sqrt{S}$ and $\phi(0)=1/\sqrt{2\pi}$. 

\subsection{Scale parameter for \texorpdfstring{$\Omega = \mathbb R$}{OmR}}

This is the same kind of density as in \eqref{pScale}, with $d=1$ and $\theta=\sigma >0$, so that $\Theta = \mathbb R^+$. But $g$ is now a symmetric density defined on the whole real line. A typical example is the class of double exponential or Laplace distributions ($g(x)=e^{-\vert x\vert}/2$), the class of symmetric normal distributions ($g(x)=\phi(x)$) or the class of symmetric Cauchy distributions ($g(x)=1/\left[\pi\left(1+x^2\right)\right]$).  It can be seen that 
\begin{align}\label{A:FBound3}
	F(I;\sigma) \approx 2\epsilon \frac{\vert x_\text{obs}\vert}{\sigma} g\left(  \frac{x_\text{obs}}{\sigma} \right),
\end{align}
whereas \eqref{PmaxScale} still holds.

\subsection{Location parameter for \texorpdfstring{$\Omega = \mathbb R$}{OmRe}}

A location parameter corresponds to $d=1$, $\theta=\mu$, $\Theta=\mathbb R$, and 
\begin{equation}
	f(x;\mu) = g(x-\mu).
	\label{pLoc}
\end{equation}
Typical examples are the family of normal distributions with fixed variance $\sigma^2$ ($g(x)=\phi(x/\sigma)/\sigma$), and the family of shifted Cauchy distributions with a fixed scale $\sigma$ ($g(x)=1/\left[\sigma \pi \left(1+(x/\sigma)^2\right)\right]$). Since 
\begin{align}\label{A:FBound4}
	F(I;\mu) \approx 2\epsilon g(x_\text{obs} - \mu), 
\end{align}
it follows that
\begin{align}
		\text{TP}_{\text{max}}(I) &\le 2\epsilon \max_{x\in\mathbb R} g(x)\\
		& = 2\epsilon C_3.
\end{align}

\subsection{Location and scale parameters for \texorpdfstring{$\Omega = \mathbb R$}{OmeRe}}\label{AS:LSR}
The two-parameter location-scale family corresponds to $d=2$, $\theta=(\mu,\sigma)$,
\begin{align*}
	f(x;\mu,\sigma) = \frac{1}{\sigma} g\left(\frac{x-\mu}{\sigma}\right),
\end{align*}
and 
\begin{equation}
	\Theta = \left\{\theta=(\mu,\sigma): \mu\in\mathbb R, \sigma>0, \frac{\mu^2}{\sigma^2}\le S\right\}.
	\label{cPDef}
\end{equation}
If the first two moments of the prior exist and $X\sim g$ is standardized to have $E(X)=0$ and $\text{Var}(X)=1$, then \eqref{cPDef} consists of all densities with a signal-to-noise ratio 
\begin{equation}
	\frac{\mu^2}{\sigma^2} \le S
\label{SNR}
\end{equation} 
that is upper bounded by a pre-chosen number $S$, as in \eqref{SNR0}. A typical example is the family of normal distributions ($g(x)=\phi(x)$). On the other hand, if $g$ corresponds to a Cauchy distribution ($g(x)=1/\left[\pi \left(1+x^2\right)\right]$) the first two moments of $X\sim g$ do not exist. Then $\mu^2/\sigma^2$ represents a generalized signal-to-noise ratio, which according to \eqref{SNR} is upper bounded by $S$.  Below we prove that  
\begin{align}
		\text{TP}_{\text{max}} (I) &\le 2\epsilon \left[\sqrt{S} \left(\max_{x\in\mathbb R}g(x)\right) + \max_{x\in\mathbb R} \{\vert x\vert g(x)\}\right]\label{PmaxBound}\\
		&=  2\epsilon \left(C_3\sqrt S + C_1\right), \label{PmaxBound2}
\end{align}
which holds whenever \eqref{PmaxCond} is satisfied, which is also a condition for the upper bound of the tuning probability to be small. In the transition from \eqref{PmaxBound} to \eqref{PmaxBound2} we assumed that $g$ is symmetric, so that $C_1=\max_{x>0} \{x g(x)\}$, as in Section \ref{AS:SPR+}. 

{\bf Proof of \eqref{PmaxBound}.} By a change of variables, it is easily seen that
\begin{align*}
	F(I;\mu,\sigma) = F\left(\frac{I}{x_\text{obs}};\frac{\mu}{x_\text{obs}},\frac{\sigma}{x_\text{obs}}\right),
\end{align*}
from which it follows that \eqref{PmaxIInv} holds, and we may, without loss of generality, assume $x_\text{obs}=1$. This gives
\begin{align}\label{A:FBound5}
\begin{aligned}
	F(I;\mu,\sigma) &= \frac{1}{\sigma} \int_{1-\epsilon}^{1+\epsilon} g\left(\frac{x-\mu}{\sigma}\right) dx\\
	&= \int_{(1-\epsilon)/\sigma}^{(1+\epsilon)/\sigma} g\left(z-\frac{\mu}{\sigma}\right)dz\\
	&\approx  \frac{2\epsilon}{\sigma} g\left(\frac{1-\mu}{\sigma}\right),
\end{aligned}
\end{align}
where in the second step we substituted $z=x/\sigma$. We have that 
\begin{equation}
	\text{TP}_{\text{max}} (I) = \max_{\mu\in\mathbb R} \text{TP}_{\text{max}}(I;\mu),
	\label{PmaxI2}
\end{equation}
where  
\begin{align}\label{Pmaxmu}
		\text{TP}_{\text{max}}(I;\mu) &= \max_{\sigma \ge \vert \mu \vert /\sqrt S} F(I;\mu,\sigma)\nonumber\\
		&\approx  \max_{\sigma \ge \vert \mu \vert/\sqrt{S}} \left\{\frac{2\epsilon}{\sigma} g\left(\frac{1-\mu}{\sigma}\right)\right\}\\
		&= 2\epsilon \cdot h\left(1-\mu,\frac{\vert \mu\vert}{\sqrt{S}}\right)\nonumber
\end{align}
whenever $\mu\ne 0$, and 
\begin{align}\label{h}
\begin{aligned}
	h(x,\sigma_0) &= \max_{\sigma\ge \sigma_0} \left\{\frac{1}{\sigma} g\left(\frac{x}{\sigma}\right)\right\}\\
	&= \max_{0<y\le \vert x\vert/\sigma_0} \left\{\frac{y}{\vert x\vert} g(y\, \mbox{sgn}(x))\right\}\\
	&= \frac{1}{\vert x\vert} \left( \max_{0<y\le \vert x\vert/\sigma_0} \{y g(y\, \mbox{sgn}(x))\}\right)\\
	&\le  \min\left\{ \frac{C_3}{\sigma_0},\frac{C_1}{\vert x\vert}\right\}.
\end{aligned}
\end{align}
Insertion of \eqref{h} into \eqref{Pmaxmu} yields
\begin{equation}
	\text{TP}_{\text{max}} (I;\mu) \le 2\epsilon \min\left\{ \frac{C_3 \sqrt{S}}{\vert \mu\vert},\frac{C_1}{\vert 1-\mu\vert}\right\}.
\label{Pmaxmu2}
\end{equation}
Finally, inserting \eqref{Pmaxmu2} into \eqref{PmaxI2}, we find that 
\begin{align}\label{PmaxIFinal}
	\text{TP}_{\text{max}}(I) &\le 2\epsilon \max_{\mu\in \mathbb R} \left\{ \min\left\{ \frac{C_3 \sqrt{S}}{\vert \mu\vert},\frac{C_1}{\vert 1-\mu\vert}\right\}\right\}\\
					&= 2\epsilon (C_3 \sqrt{S} + C_1),
\end{align}
which proves the desired result. In the last step of \eqref{PmaxIFinal} we used that the maximum of \eqref{PmaxIFinal} is obtained when $\mu$ is such that the two functions within the minimum operator have the same value. 

\subsection{Summary of results}\label{A:Summary}

The upper bounds of the tuning probability are summarized in the following table:

\begin{table}[ht!]
\centering
\caption{Maximal tuning probabilities %$\mbox{TP}_\text{max}$ of a life-permitting interval $I_X$, defined in (\ref{IXApp}), 
for diverse parametric families $\cal F$ of prior distributions, given certain constraints. %on $I_X$ and ${\cal F}$.
\label{T:Tuning}}
\resizebox{\textwidth}{!}{
\begin{tabular}{|c|c|c|c|c|c|}
\hline
$\Omega$ & ${\cal F}$ & $f(x;\theta)$ & $\Theta$ & Constraint &  $\text{TP}_{\text{max}}$  \\
\hline
& Scale & $g(x/\theta)/\theta$ & $\mathbb R^+$ & $\epsilon \ll 1$ & $2\epsilon C_1$ \\
$\mathbb R^+$ & Form and scale & $g(x/\theta_2;\theta_1)/\theta_2$ & $\mathbb R^+\times \mathbb R^+$ & None & 1\\
& Form and scale & $g(x/\theta_2;\theta_1)/\theta_2$ & $\mathbb R^+\times \mathbb R^+$ & $\text{SNR} \le S$, $\epsilon \ll 1$, $\epsilon\sqrt{S} \ll 1$, $S \gg 1$ & $2\epsilon C_2\sqrt{S}$ \\
\hline
 & Scale & $g(x/\theta)/\theta$ & $\mathbb R^+$ & $0\notin I_X$, $\epsilon \ll 1$ & $2\epsilon C_1$\\
& Scale & $g(x/\theta)/\theta$ & $\mathbb R^+$ & $0\in I_X$ & 1 \\
& Location & $g(x-\theta)$ & $\mathbb R$ & $C_3 \ll 1/\epsilon$, $\epsilon \ll 1$ & $2\epsilon C_3$\\
$\mathbb R$ & Location & $g(x-\theta)$ & $\mathbb R$ & None & 1\\
& Location and scale & $g((x-\theta_1)/\theta_2)/\theta_2$ & $\mathbb R\times \mathbb R^+$ & $\text{SNR} \le S$, $\epsilon \ll 1$, $\epsilon\sqrt{S} \ll 1$ & $2\epsilon (C_3\sqrt{S} + C_1)$\\
& Location and scale & $g((x-\theta_1)/\theta_2)/\theta_2$ & $\mathbb R\times \mathbb R^+$ & None & 1\\
\hline
\end{tabular}
}
\end{table}

The constants that appear in the rightmost column of the table are 
\begin{align}\label{A:Constants}
\begin{aligned}
	C_1 &= \max_{x>0}\{xg(x)\},\\
	C_2 &= 1/\sqrt{2\pi},\\
	C_3 &= \max_{x\in\mathbb R}g(x).
\end{aligned}
\end{align}

\begin{remark}\label{A:epsilon}
	Each proof in this section is based on the supposition $\epsilon \ll 1$. This can be seen in \eqref{A:FBound1}, \eqref{PApprox}, \eqref{A:FBound3}, \eqref{A:FBound4}, and \eqref{A:FBound5}, since in all these equations the assumption was that $\epsilon$ was small enough to warrant that the prior density of $X$ was constant over the life-permitting interval $I_X$. 
\end{remark}

\begin{remark}
	The cases where $\text{TP}_{\text{max}} = 1$ in Table \ref{T:Tuning} are produced because it is possible that the distribution is highly concentrated inside the life-permitting interval $I_X$. For instance, this happens for the location-scale parameter of Appendix \ref{AS:LSR} when the scale parameter $\sigma$ converges to zero (or $S\to\infty$). 
\end{remark}

\begin{remark}
	Since fine-tuning requires a non-degenerate random variable $X$, its variance must be positive. Therefore, suppose there exists $\sigma_0>0$ such that $\text{Var}(X) \ge \sigma_0^2$. This assumption implies that SNR$=E^2(X)/\text{Var}(X) \neq 0$ when $E(X)\neq 0$ in Table \ref{T:Tuning}. For this scenario a sufficient condition for fine-tuning is $\epsilon/\sigma_0 \ll 1$, regardless of $\mbox{SNR}$. However, the requirement $\text{Var}(X)\ge \sigma_0^2$ is not invariant with respect to scaling of $X$, and therefore less general than the dimensionless constraint $\mbox{SNR}\le S<\infty$ (which also excludes degenerate priors). Even when the 		variance does not exist (as in the Cauchy distribution), the fact that a non-degenerate random variable is under scrutiny warrants that the scale parameter $\sigma$ must be positive ($\sigma \ge \sigma_0$), and then the analogous sufficient condition for fine-tuning applies.
\end{remark}



\begin{thebibliography}{59}
\ifx \bisbn   \undefined \def \bisbn  #1{ISBN #1}\fi
\ifx \binits  \undefined \def \binits#1{#1}\fi
\ifx \bauthor  \undefined \def \bauthor#1{#1}\fi
\ifx \batitle  \undefined \def \batitle#1{#1}\fi
\ifx \bjtitle  \undefined \def \bjtitle#1{#1}\fi
\ifx \bvolume  \undefined \def \bvolume#1{\textbf{#1}}\fi
\ifx \byear  \undefined \def \byear#1{#1}\fi
\ifx \bissue  \undefined \def \bissue#1{#1}\fi
\ifx \bfpage  \undefined \def \bfpage#1{#1}\fi
\ifx \blpage  \undefined \def \blpage #1{#1}\fi
\ifx \burl  \undefined \def \burl#1{\textsf{#1}}\fi
\ifx \doiurl  \undefined \def \doiurl#1{\url{https://doi.org/#1}}\fi
\ifx \betal  \undefined \def \betal{\textit{et al.}}\fi
\ifx \binstitute  \undefined \def \binstitute#1{#1}\fi
\ifx \binstitutionaled  \undefined \def \binstitutionaled#1{#1}\fi
\ifx \bctitle  \undefined \def \bctitle#1{#1}\fi
\ifx \beditor  \undefined \def \beditor#1{#1}\fi
\ifx \bpublisher  \undefined \def \bpublisher#1{#1}\fi
\ifx \bbtitle  \undefined \def \bbtitle#1{#1}\fi
\ifx \bedition  \undefined \def \bedition#1{#1}\fi
\ifx \bseriesno  \undefined \def \bseriesno#1{#1}\fi
\ifx \blocation  \undefined \def \blocation#1{#1}\fi
\ifx \bsertitle  \undefined \def \bsertitle#1{#1}\fi
\ifx \bsnm \undefined \def \bsnm#1{#1}\fi
\ifx \bsuffix \undefined \def \bsuffix#1{#1}\fi
\ifx \bparticle \undefined \def \bparticle#1{#1}\fi
\ifx \barticle \undefined \def \barticle#1{#1}\fi
\bibcommenthead
\ifx \bconfdate \undefined \def \bconfdate #1{#1}\fi
\ifx \botherref \undefined \def \botherref #1{#1}\fi
\ifx \url \undefined \def \url#1{\textsf{#1}}\fi
\ifx \bchapter \undefined \def \bchapter#1{#1}\fi
\ifx \bbook \undefined \def \bbook#1{#1}\fi
\ifx \bcomment \undefined \def \bcomment#1{#1}\fi
\ifx \oauthor \undefined \def \oauthor#1{#1}\fi
\ifx \citeauthoryear \undefined \def \citeauthoryear#1{#1}\fi
\ifx \endbibitem  \undefined \def \endbibitem {}\fi
\ifx \bconflocation  \undefined \def \bconflocation#1{#1}\fi
\ifx \arxivurl  \undefined \def \arxivurl#1{\textsf{#1}}\fi
\csname PreBibitemsHook\endcsname

%%% 1
\bibitem{LewisBarnes2016}
\begin{bbook}
\bauthor{\bsnm{Lewis}, \binits{G.F.}},
\bauthor{\bsnm{Barnes}, \binits{L.A.}}:
\bbtitle{A {F}ortunate {U}niverse: {L}ife {I}n a {F}inely {T}uned {C}osmos}.
\bpublisher{Cambridge University Press} \blocation{}
(\byear{2016}).
\doiurl{10.1017/9781316661413}
\end{bbook}
\endbibitem

%%% 2
\bibitem{CarrRees1979}
\begin{barticle}
\bauthor{\bsnm{Carr}, \binits{B.}},
\bauthor{\bsnm{Rees}, \binits{M.J.}}:
\batitle{The anthropic principle and the structure of the physical world}.
\bjtitle{Nature}
\bvolume{278},
\bfpage{605}--\blpage{612}
(\byear{1979}).
\doiurl{10.1038/278605a0}
\end{barticle}
\endbibitem

%%% 3
\bibitem{Adams2019}
\begin{barticle}
\bauthor{\bsnm{Adams}, \binits{F.C.}}:
\batitle{The degree of fine-tuning in our universe ---and others}.
\bjtitle{Physics Reports}
\bvolume{807}(\bissue{15}),
\bfpage{1}--\blpage{111}
(\byear{2019}).
\doiurl{10.1016/j.physrep.2019.02.001}
\end{barticle}
\endbibitem

%%% 4
\bibitem{McGrewMcGrew2005}
\begin{barticle}
\bauthor{\bsnm{McGrew}, \binits{L.}},
\bauthor{\bsnm{McGrew}, \binits{T.}}:
\batitle{{On the Rational Reconstruction of the Fine-Tuning Argument}}.
\bjtitle{Philosophia Christi}
\bvolume{7}(\bissue{2}),
\bfpage{423}--\blpage{441}
(\byear{2005}).
\doiurl{10.5840/pc20057235}
\end{barticle}
\endbibitem

%%% 5
\bibitem{Barnes2017}
\begin{bchapter}
\bauthor{\bsnm{Barnes}, \binits{L.A.}}:
\bctitle{{Testing the Multiverse: Bayes, Fine-Tuning and Typicality}}.
In: \beditor{\bsnm{Chamcham}, \binits{K.}},
\beditor{\bsnm{Silk}, \binits{J.}},
\beditor{\bsnm{Barrow}, \binits{J.D.}},
\beditor{\bsnm{Saunders}, \binits{S.}} (eds.)
\bbtitle{The Philosophy of Cosmology},
pp. \bfpage{447}--\blpage{466}.
\bpublisher{Cambridge University Press} \blocation{}
(\byear{2017}).
\doiurl{10.1017/9781316535783.023}
\end{bchapter}
\endbibitem

%%% 6
\bibitem{Barnes2018}
\begin{barticle}
\bauthor{\bsnm{Barnes}, \binits{L.A.}}:
\batitle{{Fine-tuning in the context of Bayesian theory testing}}.
\bjtitle{European Journal for Philosophy of Science}
\bvolume{8}(\bissue{2}),
\bfpage{253}--\blpage{269}
(\byear{2018}).
\doiurl{10.1007/s13194-017-0184-2}
\end{barticle}
\endbibitem

%%% 7
\bibitem{Barnes2020}
\begin{barticle}
\bauthor{\bsnm{Barnes}, \binits{L.A.}}:
\batitle{{A Reasonable Little Question: A Formulation of the Fine-Tuning
  Argument}}.
\bjtitle{Ergo}
\bvolume{6}(\bissue{42}),
\bfpage{1220}--\blpage{1257}
(\byear{2019-2020}).
\doiurl{10.3998/ergo.12405314.0006.042}
\end{barticle}
\endbibitem

%%% 8
\bibitem{Collins2012}
\begin{bchapter}
\bauthor{\bsnm{Collins}, \binits{R.}}:
\bctitle{{The Teleological Argument: An Exploration of the Fine-Tuning of the
  Universe}}.
In: \beditor{\bsnm{Craig}, \binits{W.L.}},
\beditor{\bsnm{Moreland}, \binits{J.P.}} (eds.)
\bbtitle{Blackwell Companion to Natural Theology},
pp. \bfpage{202}--\blpage{281}.
\bpublisher{Wiley-Blackwell} \blocation{}
(\byear{2012}).
\doiurl{10.1002/9781444308334.ch4}
\end{bchapter}
\endbibitem

%%% 9
\bibitem{TegmarkEtAl2006}
\begin{barticle}
\bauthor{\bsnm{Tegmark}, \binits{M.}},
\bauthor{\bsnm{Aguirre}, \binits{A.}},
\bauthor{\bsnm{Rees}, \binits{M.}},
\bauthor{\bsnm{Wilczek}, \binits{F.}}:
\batitle{Dimensionless constants, cosmology, and other dark matters}.
\bjtitle{Physical Review D}
\bvolume{73}(\bissue{2}),
\bfpage{023505}
(\byear{2006}).
\doiurl{10.1103/PhysRevD.73.023505}
\end{barticle}
\endbibitem

%%% 10
\bibitem{McGrewMcGrewVestrup2001}
\begin{barticle}
\bauthor{\bsnm{McGrew}, \binits{T.}},
\bauthor{\bsnm{McGrew}, \binits{L.}},
\bauthor{\bsnm{Vestrup}, \binits{E.}}:
\batitle{{Probabilities and the Fine-Tuning Argument: A Sceptical View}}.
\bjtitle{Mind, New Series}
\bvolume{110}(\bissue{440}),
\bfpage{1027}--\blpage{1037}
(\byear{2001}).
\doiurl{10.1093/mind/110.440.1027}
\end{barticle}
\endbibitem

%%% 11
\bibitem{ColyvanGarfieldPriest2005}
\begin{barticle}
\bauthor{\bsnm{Colyvan}, \binits{M.}},
\bauthor{\bsnm{Garfield}, \binits{J.L.}},
\bauthor{\bsnm{Priest}, \binits{G.}}:
\batitle{{Problems With the Argument From Fine Tuning}}.
\bjtitle{Synthese}
\bvolume{145}(\bissue{3}),
\bfpage{325}--\blpage{338}
(\byear{2005}).
\doiurl{10.1007/s11229-005-6195-0}
\end{barticle}
\endbibitem

%%% 12
\bibitem{Kolmogorov2018}
\begin{bbook}
\bauthor{\bsnm{Kolmogorov}, \binits{A.N.}}:
\bbtitle{Foundations of the Theory of Probability},
\bedition{Second english edition} edn.
\bpublisher{Dover} \blocation{}
(\byear{2018})
\end{bbook}
\endbibitem

%%% 13
\bibitem{Bernoulli1713}
\begin{bbook}
\bauthor{\bsnm{Bernoulli}, \binits{J.}}:
\bbtitle{Ars {C}onjectandi}.
\bpublisher{Thurneysen Brothers} \blocation{}
(\byear{1713})
\end{bbook}
\endbibitem

%%% 14
\bibitem{WolpertMacReady1995}
\begin{botherref}
\oauthor{\bsnm{Wolpert}, \binits{D.H.}},
\oauthor{\bsnm{MacReady}, \binits{W.G.}}:
{No Free Lunch Theorems for Search}.
Technical Report SFI-TR-95-02-010,
Santa Fe Institute
(1995)
\end{botherref}
\endbibitem

%%% 15
\bibitem{WolpertMacReady1997}
\begin{barticle}
\bauthor{\bsnm{Wolpert}, \binits{D.H.}},
\bauthor{\bsnm{MacReady}, \binits{W.G.}}:
\batitle{{No Free Lunch Theorems for Optimization}}.
\bjtitle{IEEE Transactions on Evolutionary Computation}
\bvolume{1}(\bissue{1}),
\bfpage{67}--\blpage{82}
(\byear{1997}).
\doiurl{10.1109/4235.585893}
\end{barticle}
\endbibitem

%%% 16
\bibitem{Jaynes1968}
\begin{barticle}
\bauthor{\bsnm{Jaynes}, \binits{E.T.}}:
\batitle{{Prior Probabilities}}.
\bjtitle{IEEE Transactions On Systems Science and Cybernetics}
\bvolume{4}(\bissue{3}),
\bfpage{227}--\blpage{241}
(\byear{1968}).
\doiurl{10.1109/TSSC.1968.300117}
\end{barticle}
\endbibitem

%%% 17
\bibitem{Jaynes1957a}
\begin{barticle}
\bauthor{\bsnm{Jaynes}, \binits{E.T.}}:
\batitle{{Information Theory and Statistical Mechanics}}.
\bjtitle{Physical Review}
\bvolume{106}(\bissue{4}),
\bfpage{620}--\blpage{630}
(\byear{1957}).
\doiurl{10.1103/PhysRev.106.620}
\end{barticle}
\endbibitem

%%% 18
\bibitem{Jaynes1957b}
\begin{barticle}
\bauthor{\bsnm{Jaynes}, \binits{E.T.}}:
\batitle{{Information} {Theory} and {Statistical} {Mechanics} {II}}.
\bjtitle{Physical Review}
\bvolume{108}(\bissue{2}),
\bfpage{171}--\blpage{190}
(\byear{1957}).
\doiurl{10.1103/PhysRev.108.171}
\end{barticle}
\endbibitem

%%% 19
\bibitem{Dembski1990}
\begin{barticle}
\bauthor{\bsnm{Dembski}, \binits{W.A.}}:
\batitle{Uniform probability}.
\bjtitle{Journal of Theoretical Probability}
\bvolume{3}(\bissue{4}),
\bfpage{611}--\blpage{626}
(\byear{1990}).
\doiurl{10.1007/BF01046100}
\end{barticle}
\endbibitem

%%% 20
\bibitem{Billingsley1999}
\begin{bbook}
\bauthor{\bsnm{Billingsley}, \binits{P.}}:
\bbtitle{Convergence of Probability Measures},
\bedition{2nd} edn.
\bpublisher{Wiley} \blocation{}
(\byear{1999})
\end{bbook}
\endbibitem

%%% 21
\bibitem{DembskiMarks2009b}
\begin{barticle}
\bauthor{\bsnm{Dembski}, \binits{W.A.}},
\bauthor{\bsnm{{Marks II}}, \binits{R.J.}}:
\batitle{{Conservation of Information in Search: Measuring the Cost of
  Success}}.
\bjtitle{IEEE Transactions Systems, Man, and Cybernetics - Part A: Systems and
  Humans}
\bvolume{5}(\bissue{5}),
\bfpage{1051}--\blpage{1061}
(\byear{2009}).
\doiurl{10.1109/TSMCA.2009.2025027}
\end{barticle}
\endbibitem

%%% 22
\bibitem{DiazMarks2020a}
\begin{barticle}
\bauthor{\bsnm{D\'iaz-Pach{\'o}n}, \binits{D.A.}},
\bauthor{\bsnm{{Marks II}}, \binits{R.J.}}:
\batitle{{Generalized active information: Extensions to unbounded domains}}.
\bjtitle{BIO-Complexity}
\bvolume{2020}(\bissue{3}),
\bfpage{1}--\blpage{6}
(\byear{2020}).
\doiurl{10.5048/BIO-C.2020.3}
\end{barticle}
\endbibitem

%%% 23
\bibitem{ParkBera2009}
\begin{barticle}
\bauthor{\bsnm{Park}, \binits{S.Y.}},
\bauthor{\bsnm{Bera}, \binits{A.K.}}:
\batitle{Maximum entropy autoregressive conditional heteroskedasticity model}.
\bjtitle{Journal of Econometrics}
\bvolume{150},
\bfpage{219}--\blpage{230}
(\byear{2009}).
\doiurl{10.1016/j.jeconom.2008.12.014}
\end{barticle}
\endbibitem

%%% 24
\bibitem{ArsteinEtAl2004}
\begin{barticle}
\bauthor{\bsnm{Arstein}, \binits{S.}},
\bauthor{\bsnm{Ball}, \binits{K.}},
\bauthor{\bsnm{Barthe}, \binits{F.}},
\bauthor{\bsnm{Naor}, \binits{A.}}:
\batitle{{Solution of Shannon's Problem on the Monotonicity of Entropy}}.
\bjtitle{Journal of the American Mathematical Society}
\bvolume{17},
\bfpage{975}--\blpage{982}
(\byear{2004}).
\doiurl{10.1090/S0894-0347-04-00459-X}
\end{barticle}
\endbibitem

%%% 25
\bibitem{Barron1986}
\begin{barticle}
\bauthor{\bsnm{Barron}, \binits{A.R.}}:
\batitle{{Entropy and the Central Limit Theorem}}.
\bjtitle{The Annals of Probability}
\bvolume{14}(\bissue{1}),
\bfpage{336}--\blpage{342}
(\byear{1986}).
\doiurl{10.1214/aop/1176992632}
\end{barticle}
\endbibitem

%%% 26
\bibitem{XueEtAl2020}
\begin{barticle}
\bauthor{\bsnm{Xue}, \binits{C.}},
\bauthor{\bsnm{Liu}, \binits{J.-P.}},
\bauthor{\bsnm{Li}, \binits{Q.}},
\bauthor{\bsnm{Wu}, \binits{J.-F.}},
\bauthor{\bsnm{Yang}, \binits{S.-Q.}},
\bauthor{\bsnm{Liu}, \binits{Q.}},
\bauthor{\bsnm{Shao}, \binits{C.-G.}},
\bauthor{\bsnm{Tu}, \binits{L.-C.}},
\bauthor{\bsnm{Hu}, \binits{Z.-K.}},
\bauthor{\bsnm{Luo}, \binits{J.}}:
\batitle{{Precision measurement of the Newtonian gravitational constant}}.
\bjtitle{National Science Review}
\bvolume{7}(\bissue{12}),
\bfpage{1803}--\blpage{1817}
(\byear{2020}).
\doiurl{10.1093/nsr/nwaa165}
\end{barticle}
\endbibitem

%%% 27
\bibitem{Davies1982}
\begin{bbook}
\bauthor{\bsnm{Davies}, \binits{P.}}:
\bbtitle{The {A}ccidental {U}niverse}.
\bpublisher{Cambridge University Press} \blocation{}
(\byear{1982})
\end{bbook}
\endbibitem

%%% 28
\bibitem{DiazHossjerMarks2021}
\begin{barticle}
\bauthor{\bsnm{D\'iaz-Pach{\'o}n}, \binits{D.A.}},
\bauthor{\bsnm{H{\"o}ssjer}, \binits{O.}},
\bauthor{\bsnm{{Marks II}}, \binits{R.J.}}:
\batitle{{Is Cosmological Tuning Fine or Coarse?}}
\bjtitle{Journal of Cosmology and Astroparticle Physics}
\bvolume{2021}(\bissue{07}),
\bfpage{020}
(\byear{2021}).
\doiurl{10.1088/1475-7516/2021/07/020}
\end{barticle}
\endbibitem

%%% 29
\bibitem{Jaynes2003}
\begin{bbook}
\bauthor{\bsnm{Jaynes}, \binits{E.T.}}:
\bbtitle{Probability Theory: The Logic of Science}.
\bpublisher{Cambridge University Press} \blocation{}
(\byear{2003}).
\doiurl{10.1017/CBO9780511790423}
\end{bbook}
\endbibitem

%%% 30
\bibitem{AzharLoeb2018}
\begin{barticle}
\bauthor{\bsnm{Azhar}, \binits{F.}},
\bauthor{\bsnm{Loeb}, \binits{A.}}:
\batitle{Gauging fine-tuning}.
\bjtitle{Physical Review D}
\bvolume{98},
\bfpage{103018}
(\byear{2018}).
\doiurl{10.1103/PhysRevD.98.103018}
\end{barticle}
\endbibitem

%%% 31
\bibitem{EllisEtAl2018}
\begin{barticle}
\bauthor{\bsnm{Ellis}, \binits{G.F.R.}},
\bauthor{\bsnm{Meissner}, \binits{K.A.}},
\bauthor{\bsnm{Hermann}, \binits{N.}}:
\batitle{The physics of infinity}.
\bjtitle{Nature Physics}
\bvolume{14},
\bfpage{770}--\blpage{772}
(\byear{2018}).
\doiurl{10.1038/s41567-018-0238-1}
\end{barticle}
\endbibitem

%%% 32
\bibitem{Grabiner1983}
\begin{barticle}
\bauthor{\bsnm{Grabiner}, \binits{J.V.}}:
\batitle{{Who Gave You the Epsilon? Cauchy and the Origins of Rigorous
  Calculus}}.
\bjtitle{American Mathematical Monthly}
\bvolume{91},
\bfpage{185}--\blpage{194}
(\byear{1983}).
\doiurl{10.2307/2975545}
\end{barticle}
\endbibitem

%%% 33
\bibitem{deFinetti2008}
\begin{bbook}
\bauthor{\bparticle{de} \bsnm{Finetti}, \binits{B.}}:
\bbtitle{Philosophical Lectures on Probability}.
\bpublisher{Springer} \blocation{}
(\byear{2008})
\end{bbook}
\endbibitem

%%% 34
\bibitem{Feller1968}
\begin{bbook}
\bauthor{\bsnm{Feller}, \binits{W.}}:
\bbtitle{An Introduction to Probability Theory and Its Applications}
vol. \bseriesno{1},
\bedition{3rd} edn.
\bpublisher{Wiley} \blocation{}
(\byear{1968})
\end{bbook}
\endbibitem

%%% 35
\bibitem{Feller1971}
\begin{bbook}
\bauthor{\bsnm{Feller}, \binits{W.}}:
\bbtitle{An Introduction to Probability Theory and Its Applications}
vol. \bseriesno{2},
\bedition{2nd} edn.
\bpublisher{Wiley} \blocation{}
(\byear{1971})
\end{bbook}
\endbibitem

%%% 36
\bibitem{Doob1990}
\begin{bbook}
\bauthor{\bsnm{Doob}, \binits{J.L.}}:
\bbtitle{Stochastic Processes},
\bedition{Revised} edn.
\bpublisher{Wiley-Interscience} \blocation{}
(\byear{1990})
\end{bbook}
\endbibitem

%%% 37
\bibitem{Resnick2014}
\begin{bbook}
\bauthor{\bsnm{Resnick}, \binits{S.I.}}:
\bbtitle{A Probability Path}.
\bpublisher{Birkh\"auser} \blocation{}
(\byear{2014}).
\doiurl{10.1007/978-0-8176-8409-9}
\end{bbook}
\endbibitem

%%% 38
\bibitem{Vapnik1998}
\begin{bbook}
\bauthor{\bsnm{Vapnik}, \binits{V.}}:
\bbtitle{Statistical Learning Theory}.
\bpublisher{Wiley} \blocation{}
(\byear{1998})
\end{bbook}
\endbibitem

%%% 39
\bibitem{Einstein1905}
\begin{barticle}
\bauthor{\bsnm{Einstein}, \binits{A.}}:
\batitle{{\"Uber die von der molekularkinetischen Theorie der W\"arme
  geforderte Bewegung von in ruhenden Fl\"ussigkeiten suspendierten Teilchen}}.
\bjtitle{Annalen der Physik}
\bvolume{322}(\bissue{8}),
\bfpage{549}--\blpage{560}
(\byear{1905}).
\doiurl{10.1002/andp.19053220806}
\end{barticle}
\endbibitem

%%% 40
\bibitem{Popov2021}
\begin{bbook}
\bauthor{\bsnm{Popov}, \binits{S.}}:
\bbtitle{Two-Dimensional Random Walk: From Path Counting to Random
  Interlacements}.
\bpublisher{Cambridge University Press} \blocation{}
(\byear{2021}).
\doiurl{10.1017/9781108680134}
\end{bbook}
\endbibitem

%%% 41
\bibitem{MortersPeres2010}
\begin{bbook}
\bauthor{\bsnm{M\"orters}, \binits{P.}},
\bauthor{\bsnm{Peres}, \binits{Y.}}:
\bbtitle{Brownian Motion}.
\bpublisher{Cambridge University Press} \blocation{}
(\byear{2010}).
\doiurl{10.1017/CBO9780511750489}
\end{bbook}
\endbibitem

%%% 42
\bibitem{HossjerDiazRao2022}
\begin{barticle}
\bauthor{\bsnm{H\"ossjer}, \binits{O.}},
\bauthor{\bsnm{D{\'\i}az-Pach{\'o}n}, \binits{D.A.}},
\bauthor{\bsnm{Rao}, \binits{J.S.}}:
\batitle{{Active Information, Learning, and Knowledge Acquisition}}.
\bjtitle{PsyArXiv}
(\byear{2022}).
\doiurl{10.31234/osf.io/qt5kw}
\end{barticle}
\endbibitem

%%% 43
\bibitem{Carroll1895}
\begin{barticle}
\bauthor{\bsnm{Carroll}, \binits{L.}}:
\batitle{{What the Tortoise Said to Achilles}}.
\bjtitle{Mind}
\bvolume{104}(\bissue{416}),
\bfpage{691}--\blpage{693}
(\byear{1895}).
\doiurl{10.1093/mind/104.416.691}
\end{barticle}
\endbibitem

%%% 44
\bibitem{Godel1962}
\begin{bbook}
\bauthor{\bsnm{G{\"o}del}, \binits{K.}}:
\bbtitle{On Formally Undecidable Propositions of Principia Mathematica and
  Related Systems}.
\bpublisher{Basic Books} \blocation{}
(\byear{1962})
\end{bbook}
\endbibitem

%%% 45
\bibitem{Hofstadter1999}
\begin{bbook}
\bauthor{\bsnm{Hofstadter}, \binits{D.R.}}:
\bbtitle{G{\"o}del, Escher, Bach: an Ethernal Golden Braid}.
\bpublisher{Basic Books} \blocation{}
(\byear{1999})
\end{bbook}
\endbibitem

%%% 46
\bibitem{Anderson1972}
\begin{barticle}
\bauthor{\bsnm{Anderson}, \binits{P.W.}}:
\batitle{{More is Different: Broken symmetry and the nature of the hierachical
  structure of science}}.
\bjtitle{Science}
\bvolume{177}(\bissue{4047}),
\bfpage{393}--\blpage{396}
(\byear{1972}).
\doiurl{10.1126/science.177.4047.393}
\end{barticle}
\endbibitem

%%% 47
\bibitem{LaughlinPines1999}
\begin{barticle}
\bauthor{\bsnm{Laughlin}, \binits{R.B.}},
\bauthor{\bsnm{Pines}, \binits{D.}}:
\batitle{The theory of everything}.
\bjtitle{Proceedings of the National Academy of Sciences}
\bvolume{97}(\bissue{1}),
\bfpage{28}--\blpage{31}
(\byear{2000}).
\doiurl{10.1073/pnas.97.1.28}
\end{barticle}
\endbibitem

%%% 48
\bibitem{ThorvaldsenHossjer2020}
\begin{barticle}
\bauthor{\bsnm{Thorvaldsen}, \binits{S.}},
\bauthor{\bsnm{H{\"o}ssjer}, \binits{O.}}:
\batitle{Using statistical methods to model the fine-tuning of molecular
  machines and systems}.
\bjtitle{J Theor Biol}
\bvolume{501},
\bfpage{110352}
(\byear{2020}).
\doiurl{10.1016/j.jtbi.2020.110352}
\end{barticle}
\endbibitem

%%% 49
\bibitem{HossjerDiaz2022}
\begin{botherref}
\oauthor{\bsnm{H{\"o}ssjer}, \binits{O.}},
\oauthor{\bsnm{D\'iaz-Pach{\'o}n}, \binits{D.A.}}:
{Assessing and Testing Fine-Tuning by Means of Active Information}.
Submitted
(2022)
\end{botherref}
\endbibitem

%%% 50
\bibitem{HaugMarksDembski2021}
\begin{barticle}
\bauthor{\bsnm{Haug}, \binits{S.}},
\bauthor{\bsnm{{Marks II}}, \binits{R.J.}},
\bauthor{\bsnm{Dembski}, \binits{W.A.}}:
\batitle{{Exponential Contingency Explosion: Implications for Artificial
  General Intelligence}}.
\bjtitle{IEEE Transactions on Systems, Man, and Cybernetics: Systems}
\bvolume{52}(\bissue{5}),
\bfpage{2800}--\blpage{2808}
(\byear{2022}).
\doiurl{10.1109/TSMC.2021.3056669}
\end{barticle}
\endbibitem

%%% 51
\bibitem{Koperski2005}
\begin{barticle}
\bauthor{\bsnm{Koperski}, \binits{J.}}:
\batitle{{Should We Care about Fine-Tuning?}}
\bjtitle{British Journal for Philosophy of Science}
\bvolume{56}(\bissue{2}),
\bfpage{303}--\blpage{319}
(\byear{2005}).
\doiurl{10.1093/bjps/axi118}
\end{barticle}
\endbibitem

%%% 52
\bibitem{Bostrom2002}
\begin{bbook}
\bauthor{\bsnm{Bostrom}, \binits{N.}}:
\bbtitle{Anthropic Bias: Observation Selection Effects in Science and
  Philosophy}.
\bpublisher{Routledge} \blocation{}
(\byear{2002})
\end{bbook}
\endbibitem

%%% 53
\bibitem{McGrew2018}
\begin{barticle}
\bauthor{\bsnm{McGrew}, \binits{T.}}:
\batitle{{Fine-tuning and the Search for an Archimedean Point}}.
\bjtitle{Quaestiones Disputatae}
\bvolume{8}(\bissue{2}),
\bfpage{147}--\blpage{154}
(\byear{2018}).
\doiurl{10.5840/qd2018828}
\end{barticle}
\endbibitem

%%% 54
\bibitem{Barnes2012}
\begin{barticle}
\bauthor{\bsnm{Barnes}, \binits{L.A.}}:
\batitle{{The Fine Tuning of the Universe for Intelligent Life}}.
\bjtitle{Publications of the Astronomical Society of Australia}
\bvolume{29}(\bissue{4}),
\bfpage{529}--\blpage{564}
(\byear{2012}).
\doiurl{10.1071/AS12015}
\end{barticle}
\endbibitem

%%% 55
\bibitem{Rees2000}
\begin{bbook}
\bauthor{\bsnm{Rees}, \binits{M.J.}}:
\bbtitle{Just Six Numbers: The Deep Forces That Shape The Universe}.
\bpublisher{Basic Books} \blocation{}
(\byear{2000})
\end{bbook}
\endbibitem

%%% 56
\bibitem{SecrestEtAl2021}
\begin{barticle}
\bauthor{\bsnm{Secrest}, \binits{N.J.}},
\bauthor{\bparticle{von} \bsnm{Hausegger}, \binits{S.}},
\bauthor{\bsnm{Rameez}, \binits{M.}},
\bauthor{\bsnm{Mohayaee}, \binits{R.}},
\bauthor{\bsnm{Sarkar}, \binits{S.}},
\bauthor{\bsnm{Colin}, \binits{J.}}:
\batitle{{A Test of the Cosmological Principle with Quasars}}.
\bjtitle{The Astrophysical Journal Letters}
\bvolume{908}(\bissue{2}),
\bfpage{51}
(\byear{2021}).
\doiurl{10.3847/2041-8213/abdd40}
\end{barticle}
\endbibitem

%%% 57
\bibitem{Sarkar2022}
\begin{botherref}
\oauthor{\bsnm{Sarkar}, \binits{S.}}:
{Heart of Darkness}.
Inference
\textbf{6}(4)
(2022).
\doiurl{10.37282/991819.22.21}
\end{botherref}
\endbibitem

%%% 58
\bibitem{Conrad2005}
\begin{botherref}
\oauthor{\bsnm{Conrad}, \binits{K.}}:
Probability {D}istributions and {M}aximal {E}ntropy
(2005).
\url{http://www.math.uconn.edu/~kconrad/blurbs/analysis/entropypost.pdf}
\end{botherref}
\endbibitem

%%% 59
\bibitem{Billingsley1995}
\begin{bbook}
\bauthor{\bsnm{Billingsley}, \binits{P.}}:
\bbtitle{Probability and Measure},
\bedition{3rd.} edn.
\bpublisher{Wiley} \blocation{}
(\byear{1995})
\end{bbook}
\endbibitem

\end{thebibliography}
\end{document}